\documentclass[referee,a4paper,12pt,traditabstract]{swsc} 

\usepackage{graphicx}
\usepackage[final]{changes}
\usepackage{epstopdf}
\usepackage[displaymath,mathlines]{lineno}
\usepackage[authoryear,round]{natbib}
\usepackage[backref]{hyperref}
\usepackage{url}
\usepackage[utf8]{inputenc}
\usepackage{amstext}
\usepackage{booktabs}
\usepackage{multirow} 
\usepackage{longtable}

\usepackage{savesym}
\usepackage{amsmath}
\savesymbol{iint}
\usepackage{txfonts}
\restoresymbol{TXF}{iint}

\hypersetup{colorlinks=true,citecolor=cyan,urlcolor=cyan,linkcolor=blue}

\begin{document}


   \title{Probabilistic Prediction of Geomagnetic Storms and the K$_{\textrm{p}}$ Index}

   
   \titlerunning{Probabilistic Storm Forecast}

   \authorrunning{Chakraborty and Morley}

   \author{S. Chakraborty
          \inst{1}
          \and
          S. K. Morley\inst{2}
          }

   \institute{Bradley Department of Electrical \& Computer Engineering, Virginia Tech, VA, USA\\
              \email{\href{mailto:shibaji7@vt.edu}{shibaji7@vt.edu}}
         \and
             Space Science and Applications (ISR-1), Los Alamos National Laboratory, NM, USA\\
             \email{\href{mailto:smorley@lanl.gov}{smorley@lanl.gov}}
             }


 
  \abstract
       {
       Geomagnetic activity is often described using summary indices to summarize the likelihood of space weather impacts, as well as when parameterizing space weather models.
       The geomagnetic index $\text{K}_\text{p}$ in particular, is widely used for these purposes.
       Current state-of-the-art forecast models provide deterministic $\text{K}_\text{p}$ predictions using a variety of methods -- including empirically-derived functions, physics-based models, and neural networks -- but do not provide uncertainty estimates associated with the forecast.
       This paper provides a sample methodology to generate a 3-hour-ahead $\text{K}_\text{p}$ prediction with uncertainty bounds and from this provide a probabilistic geomagnetic storm forecast.
       Specifically, we have used a two-layered architecture to separately predict storm ($\text{K}_\text{p}\geq 5^-$) and non-storm cases.
       As solar wind-driven models are limited in their ability to predict the onset of transient-driven activity we also introduce a model variant using solar X-ray flux to assess whether simple models including proxies for solar activity can improve the predictions of geomagnetic storm activity with lead times longer than the L1-to-Earth propagation time.
       By comparing the performance of these models we show that including operationally-available information about solar irradiance enhances the ability of predictive models to capture the onset of geomagnetic storms and that this can be achieved while also enabling probabilistic forecasts.
    
       }        


   \maketitle

\section{Introduction}
Modern electrical systems and equipment on the Earth such as navigation, communication, satellite and power grid systems can be affected by space weather \citep[e.g.,][]{Choi2011,Qiu2015,Morley2019}. 
The societal impact of space weather is increasing \citep{Schrijver2015,Eastwood2017} and operational centers provide a range of predictions for end-users \citep{Bingham2019}, including geomagnetic storm predictions based on the Kp index \citep{Sharpe2017}.

$\text{K}_\text{p}$ is a planetary 3-hour averaged range index that describes the intensity of the magnetic disturbance \citep{Bartels1949}.
$\text{K}_\text{p}$ starts from 0 (very quiet) to 9 (very disturbed) with 28 discrete values described by $0, 0^+, 1^-, 1,1^+, ... , 9^-, 9$ \citep{Bartels1949}.
The National Oceanic and Atmospheric Administration (NOAA) Space Weather Prediction Center (SWPC) classifies geomagnetic activity in six levels, shown in Table 1, based on the ranges of $\text{K}_\text{p}$.
In addition to being a forecast product in its own right, the $\text{K}_\text{p}$ index is widely used as an input to other magnetospheric models \citep[e.g.,][]{Carbary2005ABoundaries} including some aimed at operational use \citep{OBrien2009,Horne2013}.

\begin{table}[!hbt]
\centering
\label{table:stormlevel}
\begin{tabular}{cc}\toprule
\textbf{Storm Level} & \textbf{$\text{K}_\text{p}$ Range} \\ \midrule
$\text{G}_\text{0}$& $\text{K}_\text{p}<\text{5}^\text{-}$\\ 
$\text{G}_\text{1}$& $\text{5}^\text{-}\leq \text{K}_\text{p}<\text{6}^\text{-}$\\
$\text{G}_\text{2}$& $\text{6}^\text{-}\leq \text{K}_\text{p}<\text{7}^\text{-}$\\
$\text{G}_\text{3}$& $\text{7}^\text{-}\leq \text{K}_\text{p}<\text{8}^\text{-}$\\
$\text{G}_\text{4}$& $\text{8}^\text{-}\leq \text{K}_\text{p}<\text{9}^\text{-}$\\
$\text{G}_\text{5}$& $\text{K}_\text{p}>\text{9}^\text{-}$\\ \bottomrule
\end{tabular}
\caption{Table showing different categories of geomagnetic storm and associated $\text{K}_\text{p}$ levels. The categorization is done based on the intensity of the geomagnetic storm following the NOAA SWPC scales.}
\label{tab:1}
\end{table}

There are three main categories of models for $\text{K}_\text{p}$ prediction: coupled physics-based models such as the Space Weather Modeling Framework \citep[e.g.,][]{Toth2005,Haiducek2017}; ``traditional'' empirical models \citep{newell_nearly_2007,luo_two_2017}; and machine-learning models \citep{costello_moving_1998,boberg_real_2000,wing_kp_2005,Wintoft2017ForecastingValues,tan_geomagnetic_2018,Sexton2019KpNetwork}. 

Longer lead-time predictions are typically the domain of empirical \citep[e.g.,][]{luo_two_2017} and machine-learning models \citep[e.g.,][]{Sexton2019KpNetwork}.
Since the first neural network-based $\text{K}_\text{p}$ forecasting model proposed by Costello \citep{costello_moving_1998}, many subsequent forecast models \citep{boberg_real_2000,wing_kp_2005,tan_geomagnetic_2018} implement different variants of the neural network to improve the forecast accuracy. To date, none of these predictions have provided a probabilistic prediction, and very few attempted to characterize the uncertainty associated with the predicted $\text{K}_\text{p}$ value~\citep[see, e.g.,][]{Wintoft2017ForecastingValues}.
With the development of new machine-learning techniques, recent $\text{K}_\text{p}$ and storm forecast models come with much higher accuracy, but few have separately examined the model performance under different ranges of geomagnetic storm conditions \citep[see, e.g.,][]{Zhelavskaya2019}.
Recent work on $\text{K}_\text{p}$ prediction by \citep{Shprits2019} highlighted the inherent limitation of solar wind-driven models for long lead-time predictions, and noted that ``further improvements in long‐term modeling should include ... empirical modeling driven by observations of the Sun''.

To assess the viability of moving beyond solar wind-driven models using operationally-available data, we also investigate the inclusion of solar X-ray flux data as a model parameter.
Solar X-ray flux was chosen as recent studies have shown that these data can be used to forecast solar flare activity \citep{Winter2015UsingClass} as well as Solar Energetic Particle (SEP) events \citep{Nunez2018PredictingElectrons}.
While the majority of large geomagnetic storms are caused by CMEs \citep[e.g.,][]{Gonzalez1999}, it has been shown that CMEs are correlated with solar flare activity \citep{zhou_correlation_2003, Kay2003TheCMEs, Wu2008EnergyActivity, Lippiello2008DifferentEjections}.
Further, solar X-ray flux is operationally available from the NOAA GOES satellites (see \url{https://www.swpc.noaa.gov/products/goes-x-ray-flux}).
We, therefore, include GOES X-ray data in a variant of our predictive model to determine whether its use, as a proxy of solar magnetic activity, allows us to better capture the (often CME-driven) geomagnetic storms. 

The primary aim of this paper is to generate a $\text{K}_\text{p}$ prediction with associated uncertainty bounds and exploit it to provide a probabilistic geomagnetic storm forecast.
The secondary objective is to test whether a simple, operationally-available solar irradiance data set can be included in this framework to better capture the effects of solar transients.
We use a machine-learning method to forecast $\text{K}_\text{p}$, including the associated uncertainty, then exploit the uncertainty bounds in $\text{K}_\text{p}$ to provide a probabilistic forecast.
The paper is organized as follows: Section 2 explains the data analysis and the model development; Section 3 describes the results, shows how we develop a probabilistic storm forecast and assesses the model performance; and finally we discuss the results in the context of similar work.

\section{Data Analysis \& Model Architecture}
Here we describe the data sets and the model architecture used in this study.
Specifically, we present the basic characteristics of the data sets and a brief justification of our choices of model features.
In addition, we provide a short introduction to the technical terms and metrics used to evaluate the model performance.
Finally, we describe the construction of our predictive models and note some strengths and weaknesses of our approach.

\subsection{Model Features and Data Sources}
The solar wind energy and magnetospheric coupling are known to be well-described by the solar wind parameters and the state of the magnetosphere \citep{Dungey1961, Baker1981}.
However, many of the solar wind parameters are correlated with each other and might carry redundant solar wind structure information~\citep[e.g.,][]{Hundhausen1970,Wing2016InformationBelt}.
There is a long history of using plasma moments (number density and velocity) and the interplanetary magnetic field to describe the coupling of energy from the solar wind into the magnetosphere \citep[e.g.,][]{Baker1981, borovsky_transport_1998, xu_new_2015}.
More recently-developed coupling functions using solar wind parameters have been shown to have better correlations with geomagnetic indices \citep[e.g.,][]{newell_nearly_2007} and clear physical origins \citep[e.g.,][]{Borovsky2013}.
It is also clear that including a measure of the present state of the magnetosphere helps predict the evolution of global indices like $\text{K}_\text{p}$ \citep{Borovsky_canonical_2014,luo_two_2017}.

Based, in part, on these considerations we use just nine solar wind parameters and the historical $\text{K}_\text{p}$ as model features (input parameters). The input parameters are chosen based on the studies done by the previous studies~\citep{newell_nearly_2007,Borovsky2014,xu_new_2015}. These model features are listed in Table 2 along with the notation we use in this paper and any transformations applied prior to model training.
For the solar wind data, we use 20 years of 1-minute resolution values, starting from 1995, obtained from NASA's OMNIWeb service (\url{https://omniweb.gsfc.nasa.gov/ow_min.html}).
The 3-hourly $\text{K}_\text{p}$ index is obtained from the GFZ German Research Centre for Geosciences at Potsdam (\url{https://www.gfz-potsdam.de/en/kp-index/}).

\begin{figure}[h]
 \centering
 \includegraphics[scale=0.6]{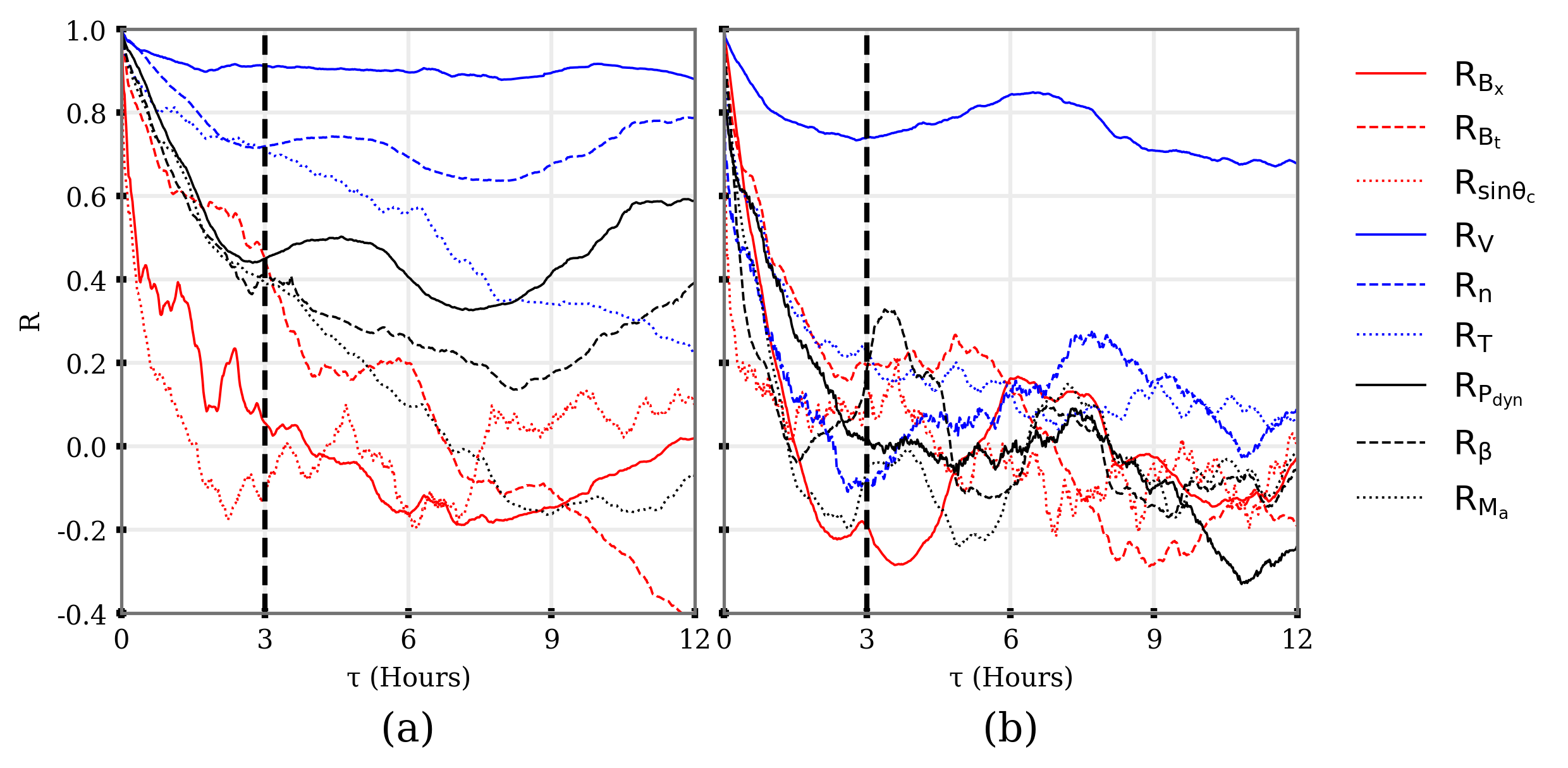}
 \caption{Auto--correlation functions of different solar-wind parameters during: (a) solar minima, and (b) solar maxima.}
 \label{figure1}
  \end{figure}

As solar wind structures are spatial in nature, the measured parameters are auto-correlated. Solar wind data from L1 monitors are point measurements and hence spatial structure along the Sun-Earth line manifests as a temporal correlation. Hence, we performed an auto-correlation analysis of all the solar wind parameters as presented in Figure~\ref{figure1}.
From the figure, we can conclude that, during solar minimum, most of the solar wind parameters are highly autocorrelated for a longer duration, while during solar maximum, the correlation coefficient drops within a few hours. 
This suggests more transients in solar wind during solar maximum than the solar minimum, consistent with observations \citep{Richardson2012}.
All parameters selected as model features display auto-correlation, and most parameters decorrelate within 3 hours, with solar wind speed being a notable exception.
Indeed, at solar maximum, all parameters except solar wind speed decorrelate in less than three hours.
At solar minimum, the auto-correlation for the majority of parameters falls below 0.5 within three hours.
We, therefore, used 3-hourly averaged solar wind data to train our models, which has the added benefit of placing the input data on the same cadence as the predicted output. \added{Note that the goal of this linear analysis (autocorrelation) is to describe the redundant information given in subsequent samples of individual solar wind parameters, rather than identifying time lags in the magnetospheric response to solar wind driving. In addition, we acknowledge that including nonlinear correlations may provide more robust estimates of the correlation scales, and could exhibit different behaviors \citep[e.g.,][]{Johnson2018,Wing2016InformationBelt}.}

As the $\text{K}_\text{p}$ index is quasi-logarithmic and essentially categorical \citep{Mayaud1980}, we converted the reported $\text{K}_\text{p}$ values to decimal numbers using $\text{k}\pm\frac{1}{3}$ following \citet{tan_geomagnetic_2018}.

\begin{table}[!hbt]
\centering
\label{table:inputfeatures}
\begin{tabular}{lcccc}\toprule
\textbf{Features} & \textbf{Symbol} & \textbf{Units} & \textbf{Transformation} & \textbf{Model} \\ \midrule
Anti-sunward component of IMF & $\text{B}_\text{x}$ & nT & Box-Cox & All\\ 
Transverse component of IMF & $\text{B}_\text{T}=\sqrt{\text{B}^\text{2}_\text{y}+\text{B}^\text{2}_\text{z}}$ & nT & Box-Cox & All\\
IMF clock angle & $\theta_\text{c}=\tan^{-1}\left({\frac{\text{B}_\text{y}}{\text{B}_\text{z}}}\right)$ & radian & Box-Cox & All\\ 
Bulk Velocity & $\text{v}=\sqrt{\text{v}_\text{x}^\text{2}+\text{v}_\text{y}^\text{2}+\text{v}_\text{z}^\text{2}}$ & $\text{km}$ $\text{s}^\text{-1}$ & Box-Cox & All\\
Number Density & n & $\text{m}^\text{-3}$ & Box-Cox & All\\
Electron Temperature & T & K & Box-Cox & All\\
Dynamic Pressure& $\text{P}_\text{dyn}$ & $\text{Nm}^\text{-2}$ & Box-Cox & All\\
Mach number& $\text{M}_\text{a}$ & 1 & Box-Cox & All\\
Plasma Beta& $\beta$ & 1 & Erf & All\\
Historical $\text{K}_\text{p}$& $\text{K}_\text{p}^\text{(t-3)}$ & 1 & Box-Cox & All\\
Background X-ray & B & W$\text{m}^\text{2}$& None & LSTM$_c$, $\text{dGPR}^{+}_\text{s}$\\
X-ray Flux Ratio & R & 1 & None & LSTM$_c$,$\text{dGPR}^{+}_\text{s}$\\
\bottomrule
\end{tabular}
\caption{Table showing input features of the model, transformations used (if any), and which models use each feature. The given transformation is only used prior to fitting the Gaussian process model. Box-Cox $\rightarrow$ $\text{B(x)}$=$\text{sgn(x)}\times\text{log}_\text{e}\text{x}$ and Erf$\rightarrow$ $\text{erf(x)}$=$\text{sgn(x)}\times\sqrt{\text{2}}\text{erf}^\text{-1}{\left(\text{1-2e}^{\frac{\text{a}|\text{x}|}{\text{b}}}\right)}$}.
\label{tab:3}
\end{table}

In this study, we will also test the effectiveness of including a previously-unused set of solar data for $\text{K}_\text{p}$ prediction.
That is, we will introduce our modeling approach using a standard construction with only upstream solar wind measurements to drive the model.
We will then train a second model that differs only in the inclusion of X-ray flux in the model features.

As described above, the idea behind using the solar X-ray data is that the upstream solar wind carries little to no information about transients that are moving towards Earth.
Advanced notice of transients from in-situ L1 measurements is therefore limited to periods typically less than 1 hour.
By including solar data, even a coarse measure such as the X-ray flux, we aim to demonstrate that additional information about the likelihood of transients can be included in the model and improve forecasts with a lead time longer than the L1-to-Earth propagation time.
In other words, we treat the X-ray flux data as a proxy for the magnetic complexity of the Sun and anticipate that including this data will allow the model to predict the arrival of CMEs earlier than a model-driven purely by solar wind measurements.
We obtain the GOES X-ray flux data from NOAA's National Centers for Environmental Information (\url{https://satdat.ngdc.noaa.gov/sem/goes/data/}).

The GOES X-ray sensors have 2 channels that measure solar irradiance in two wavebands, namely, hard (0.05-0.4\,nm) and soft (0.1-0.8\,nm) X-rays.
In this study, we followed \citet{Winter2015UsingClass} in using a background term and a flux ratio derived from the two GOES wavebands.
The X-ray background (B) has been computed as the smoothed minimum flux in a 24-hour window preceding each 1-minute GOES soft X-ray flux observation. In a recent study, the X-ray background parameter was found to describe the solar activity cycle better than the daily F10.7 parameter~\citep{winter_estimate_2014-1}.
The X-ray flux ratio (R) has been calculated by taking the ratio of hard X-ray flux over soft X-ray flux, and provides a measure of the temperature of the flare emission~\citep{garcia1994temperature}.
\citet{Kay2003TheCMEs} showed that flares associated with CMEs tended to have lower temperatures, at a given intensity level, than flares without CMEs. Thus both the soft X-ray flux and the flux ratio, R, are important for determining the likelihood of an eruptive event. Further, \citet{Michalek2009TwoEjections} showed a good correlation between the energy of the CME and the peak of the X-ray flare. Finally, recent studies showed that the X-ray flux ratio is a good predictor for extreme solar events~\citep{Nunez2018PredictingElectrons,Kahler2018ForecastingRatios}.

\begin{figure}[h]
 \centering
 \includegraphics[scale=0.7]{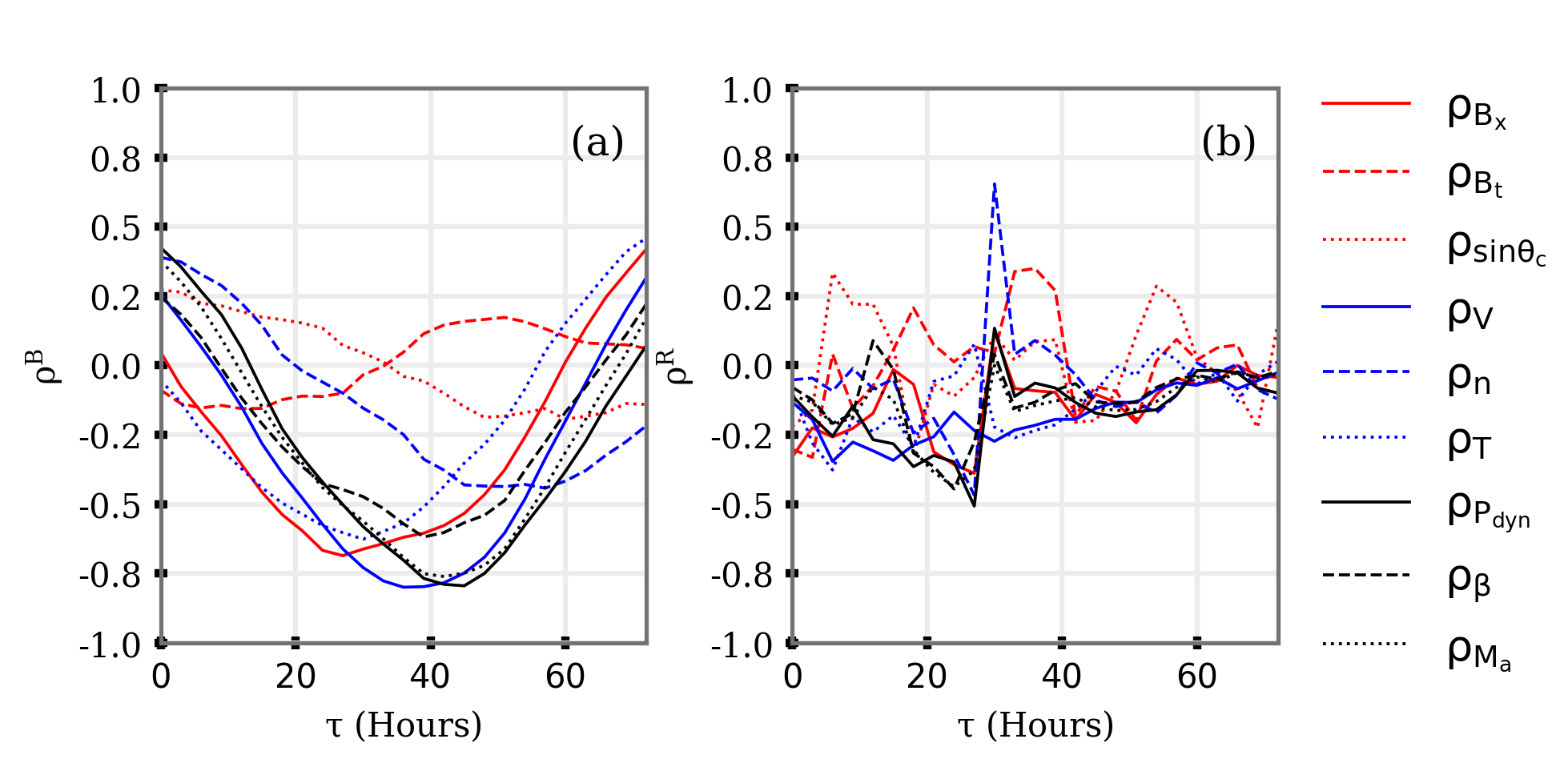}
 \caption{Cross--correlation functions of different solar-wind parameters with (a) GOES flux background (B) and (b) ratio (R) of hard (0.05--0.4 nm) and soft (0.1--0.8 nm) X--ray flux data.}
 \label{figure2}
  \end{figure}

As X-rays propagate from the Sun to the Earth at the speed of light, there will, of course, be a time lag associated with the arrival at Earth of any related geomagnetic activity due to associated coronal mass ejections.
In this preliminary work to demonstrate the utility of including solar X-ray flux data in a $\text{K}_\text{p}$ prediction model, we assume a constant time lag that we apply to the derived X-ray products B and R.
Figure~\ref{figure2} presents a time-lagged cross-correlation analysis of B and R with other solar wind parameters to highlight any time lags between these two data sets.
The correlation analysis shows that the X-ray background (B) parameter is significantly correlated with many solar wind parameters at lags around 48 hours.
The correlations between the X-ray flux ratio (R) and the solar wind parameters are smaller and less consistent across solar wind parameters. \added{A lack of clear, or strong, linear correlations with solar wind parameters at a given fixed lag does not necessarily indicate that the parameter R is confounding. Better lag estimates could be obtained using nonlinear analysis~\citep[e.g.,][]{Johnson2018,Wing2016InformationBelt}, however, models used in this study can extract nonlinear relationships. We therefore expect nonlinear relationships in the dataset to be captured by the proposed models.}

\added{Transit times of coronal mass ejections can range from less than 20 hours to more than 80 hours \citep[e.g.,][]{Schwenn2005}, however faster CMEs tend to be more geoeffective \citep{Gonzalez1999,Srivastava2002}.}
\replaced{Hence we}{We, therefore,} apply a constant time lag of 48 hours to the X-ray flux derived parameters, consistent with a typical travel time from the Sun to Earth of geomagnetic-storm associated interplanetary CMEs \citep{Srivastava2004}. Although we note that in future work it may be useful to explore the effects of choosing this lag over a different fixed lag, or even use of a variable lag.

\subsection{Technical Definitions \& Metrics for Model Evaluation}
In this subsection, we define the technical terms and the metrics that we use in the latter part of this study to evaluate and compare the models' performance. Good overviews of model performance metrics and model validation methodologies targeted at space weather have been given by \citet{MorleyMetrics2018}, \citet{Liemohn2018ModelPredictions}, and \citet{Camporeale2019}.

For binary event analysis, we define correct ``yes'' events as \textit{True Positives} (denoted as $a$). Similarly, we call incorrect ``yes'' events \textit{False Positives (b)}, incorrect ``no'' events are \textit{False Negatives (c)}, and correct ``no'' events are \textit{True Negatives (d)}.
\begin{equation*}
\left.\begin{aligned}
\text{ROC curve} &= \text{A graphical diagnostic illustration of a binary classifier.}\\
\text{AUC} &= \text{Area under the ROC curve.}\\
\text{Probability of Detection (PoD)} &= \frac{a}{a+c}\\
\text{Probability of False Detection (PoFD)} &= \frac{b}{b+d}\\ 
\text{False Alarm Ratio (FAR)}  &= \frac{b}{a+b}\\
\text{Bias}  &= \frac{a+b}{a+c}\\
\text{Critical Success Index (CSI)}  &= \frac{a}{a+b+c}\\
\text{Matthews correlation coefficient (MCC)} &= \frac{a\times d-b\times d}{\sqrt{(a+b)(a+c)(b+d)(c+d)}}\\
\text{Heidke Skill Score (HSS)}  &= \frac{2\times(ad-bc)}{(a+c)(c+d)+(a+b)(b+d)}\\
\end{aligned}\right.
\end{equation*}
Note that a perfect forecast will have HSS, MCC, PoD, Bias, CSI, $R^2$ equal to 1 and FAR equal to 0.
Unskilled or random forecasts will have HSS, MCC, PoD, CSI, $R^2$ of 0, FAR of 1.

For assessing numerical predictions of the $\text{K}_\text{p}$ index we use:
\begin{equation*}
\left.\begin{aligned}
\text{Root mean square error (RMSE) }(\epsilon) &= \sqrt{\frac{1}{\text{n}}\sum_i(x_i-\hat{x_i})^{2}}\\
    \text{Mean Absolute Error (MAE) } (\bar{|\delta|}) &= \frac{1}{n}\sum_i|x_i-\hat{x_i}|\\
    \text{Coefficient of Determination } (R^2) &= 1-\frac{\sum_i(x_i-\hat{x_i})^{2}}{\sum_i(x_i-\frac{1}{n}\sum_ix_i)^{2}}
    \end{aligned}\right.
\end{equation*}
where, $x_i$, $\hat{x_i}$ and $n$ are observations, predicted output and a total number of observations, respectively.
A perfect model will have zero RMSE and MAE, while a coefficient of determination of 1.

\subsection{Model Development}
\begin{figure}[h]
 \centering
 \includegraphics[scale=0.75]{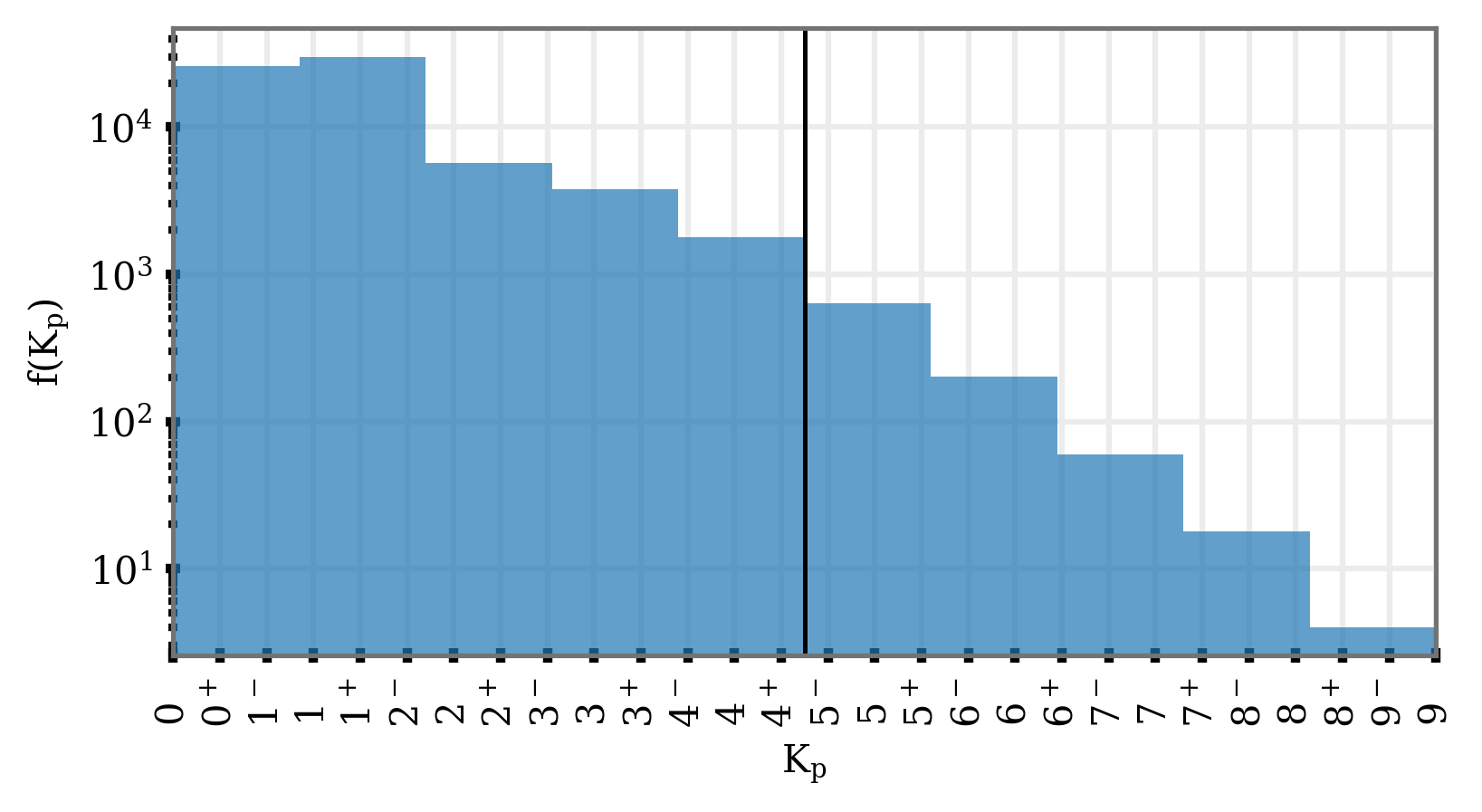}
 \caption{Distribution of $\text{K}_\text{p}$. 20 years 1995--2014 of data has been used in this plot. $f(\text{K}_\text{p})$ is the frequency (i.e., the number of events) plotted on a logarithmic scale. The black vertical line is $\text{K}_\text{p}$=$5^-$.}
 \label{figure3}
  \end{figure}

Figure~\ref{figure3} presents the distribution of 22 years of $\text{K}_\text{p}$ values, where the log-scaled frequency of occurrence is shown on the vertical axis and the $\text{K}_\text{p}$ value is shown on the horizontal axis.
From the figure, it is evident that most of the events are distributed between $[0,5^-)$, a relatively small number of events are distributed between $[5^-,7^-)$, and there are very few extreme events $\geq 7$.
Following the NOAA G-scale for geomagnetic storms, we choose $\text{K}_\text{p} \geq 5^{-}$ as the threshold for ``storm-time'', marked with a vertical line in Figure~\ref{figure3}.
Using this division, storm-time comprises approximately one-twentieth of the data set. 
This number continues to drop rapidly as $\text{K}_\text{p}$ increases.
If we take the ratio of more extreme events ($K_P\geq 8^+$) versus non-storm events, the number drops to $\approx\frac{1}{200}$.
This effect is known as data imbalance \citep[e.g.,][]{Estabrooks2004} and can lead to significant errors in a model fit to the data without accounting for the imbalance \citep[see, e.g.,][]{Shprits2019}.

Both oversampling and undersampling techniques have been used to address data imbalance \citep{Estabrooks2004,Shprits2019}, and both methods help develop models with a better predictive skill \citep{Yap2014}. 
Choosing a single regression model to predict $\text{K}_\text{p}$ across quiet and disturbed geomagnetic conditions will likely not provide an optimal forecast unless the data imbalance is addressed.
Here we take a similar approach to \citet{tan_geomagnetic_2018}, \citet{boberg_real_2000} and minimize the data imbalance by separating the problem into two different categories: storm-time and non-storm.
This then leads to the first step in our model architecture, where we use a classifier to determine quiet or active conditions, and subsequently use a probabilistic regressor to predict the $\text{K}_\text{p}$ value. 

\begin{figure}[h]
    \hspace{-2.2cm}
    \includegraphics[scale=0.72]{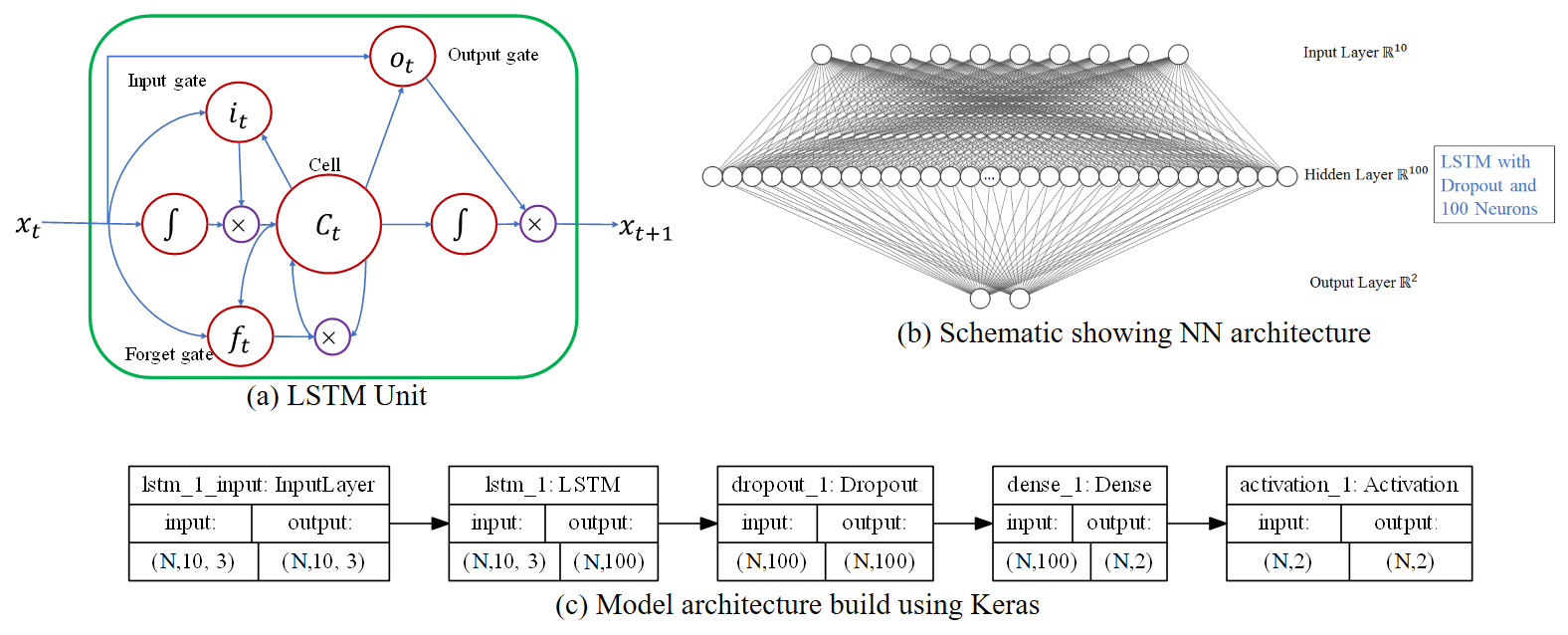}
    \caption{Schematics showing architectures (a) of a single LSTM block, (b) of the classifier consisting of one input layer, one LSTM layer consists of 100 nodes (neurons), dropout, and output layer, and (c) of the classifier model implemented using Keras.}
    \label{figure4}
\end{figure}

In this study, we have used a deterministic long-short term memory (LSTM) neural network~\citep{Sepp1997}. An LSTM is a type of recurrent neural network, where a ``memory cell'' is used to store information between time steps \citep[see][and references therein]{tan_geomagnetic_2018}. LSTM neural networks are a special type of recurrent neural network, well-suited to time series data analysis, and they require continuous data for training. \added{However, we encounter missing IMF and solar wind data values issue. To handle the missing data issue we use the interpolation method described in~\citet[see section 2.1][and references therein]{tan_geomagnetic_2018}. The method used by~\citet{tan_geomagnetic_2018} is appropriate for relatively small (up to 12 samples) data gaps, for larger data gaps we discarded the samples.}
As with other types of neural networks, an LSTM can be used as either a regressor or a classifier.
The layout of our LSTM classifier is shown schematically in Figure~\ref{figure4}. Panel \ref{figure4}a presents a schematic of a single LSTM unit. The LSTM unit consists of input, output, forget gates, and memory cell ($C_t$)~\citep[see][and references therein]{Sepp1997}. Panel \ref{figure4}b illustrates the overall architecture of the classifier, where the central layer is comprised of LSTM units. Panel \ref{figure4}c shows the implementation as layers in the Keras model. The ``N'' in the input/output shape of the model blocks shows the number of time samples, which can be varied at run-time. However, as described later in this section we use a 27-day window at 3-hourly cadence, therefore $\text{N} = 216$.

As described in the previous paragraph LSTM classifier comprises an input layer with 10 neurons, a hidden layer with 100 neurons, and an output layer of 2 neurons.
The MSE on the validation data reduced as the number of neurons in our hidden layer increased, but adding more than $\sim$100 LSTM units led to smaller improvements. We therefore chose to retain only 100 LSTM units in our hidden layer. To help reduce overfitting and increase the generalizability of the model we included dropout regularization.
The input shape of the data is therefore $(N\times10\times3)$, where $\text{N} = 216$, 10 and 3 are the number of data points, the number of input parameters (see Table 2), and the time-lagged units (the input vector at times $\text{t}$, $\text{t}-1$, and $\text{t}-2$), respectively. Hence, the input shape for one data point is $10\times3$. Note, the input shape for one data point can also be $12\times3$, based on the choice of model ($\mu^\text{OMNI}$ or $\mu^{\text{OMNI}^\text{+}}$, see the following paragraphs for details).
To ensure that the classifier performance can be generalized beyond the training data, we split available data into two categories: training/validation and testing.
For training and validation we used the intervals 1995--2000 and 2011--2014.
To mitigate the effects of data imbalance we used a random downsampling strategy to balance the storm-time and no-storm intervals.
After downsampling (from 29200 to 5716 points), we split the data into ``training'' and ``validation'' sets and train the classifier, where the validation data is used for tuning the model parameters and comprises 20\% of the training set.
Data from 2001--2010 was reserved for out-of-sample testing of the predictions (26645 points). The rationale behind using the above mentioned periods for train/validation and test the model is to increase the model \replaced{performance }{throughput} and reduce the model bias. Both train and test periods consist of different solar maximum and minimum data to capture solar cycle dynamics and testing with out-of-sample data ensures that the model generalizes well to unseen data.

\begin{figure}[h]
		\begin{center}
		\includegraphics[scale=1.0]{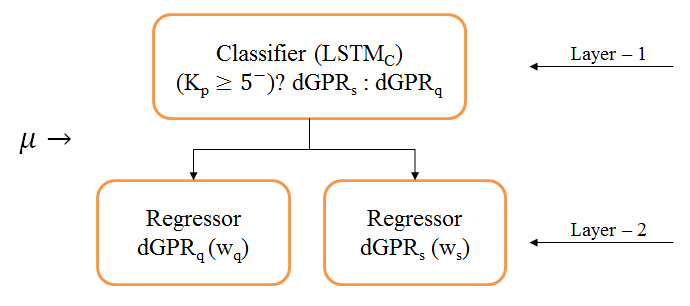}
		\caption{Proposed model ($\mu$) architecture: Classifier is deterministic in nature, and regressors are probabilistic in nature. The threshold for the classifier is $\text{K}_\text{p}$=$5^-$. Here, $w_q$ \& $w_s$ are the variable training windows for two regressors. For details refer text.}
		\end{center}
		\label{figure5}
	\end{figure}

Following \citet{tan_geomagnetic_2018} we employ a two-layer architecture, where we use separate regression models to predict $\text{K}_\text{p}$ under quiet or active geomagnetic conditions.
Unlike \citet{tan_geomagnetic_2018} we use probabilistic regressors.
The model structure is shown in Figure~\ref{figure5}.
The prediction made by the classifier is used to determine which regressor is going to be selected.
As the primary aim of this work is to produce a probabilistic prediction of $\text{K}_\text{p}$, we chose regression models that output a distribution rather than a single value. We used semi-parametric Deep Gaussian Process Regression, commonly known as deepGPR, to build the regressors.
DeepGPR (dGPR) is a Gaussian Process model with neural network architecture.
A Gaussian Process is a random process that follows a multivariate normal distribution.
Specifically, dGPR \citep{alshedivat2016srk} is a Gaussian Process Regression \citep{Rasmussen2006GaussianLearning} method, which uses a deep LSTM network to optimize the hyperparameters of the Gaussian Process kernels.

For example, if the classifier predicts a geomagnetic storm then regressor $\text{dGPR}_\text{s}$ is selected, otherwise regressor $\text{dGPR}_\text{q}$ is used.
At each time step, the dGPR is retrained using the interval of preceding data (the training window), and thus our regressors are dynamic non-linear models.
Dynamic models do not need to assume a constant functional relationship between the model covariates (e.g., solar wind drivers) and response ($\text{K}_\text{p}$). Static models implicitly combine the effects of any potentially time‐varying relationships in the error terms or average over the effects in the estimation of model coefficients \citep[see, e.g.,][]{osthus14}. By using a relatively short, local training window, the data is used more efficiently and computational complexity is reduced. For training and validation of the dGPR-based dynamic model, including training window selection, we used the mean squared error (MSE) as our loss function.

Optimizing the hyperparameters of a dGPR is much easier while working with input parameters that are normally-distributed. To ensure better behavior of the Gaussian process model we transformed all the substantially non-normally distributed input parameters listed in Table 2 using either a Box-Cox transform or the Complementary Error Function. After the transformation, we check the skewness and kurtosis of the transformed variable to validate the transformation. We found Box-Cox transformation worked well for all IMF and solar wind parameters except plasma beta. We transform the plasma beta using the Complimentary Error Function.

\begin{figure}[h]
 \centering
 \includegraphics[scale=0.75]{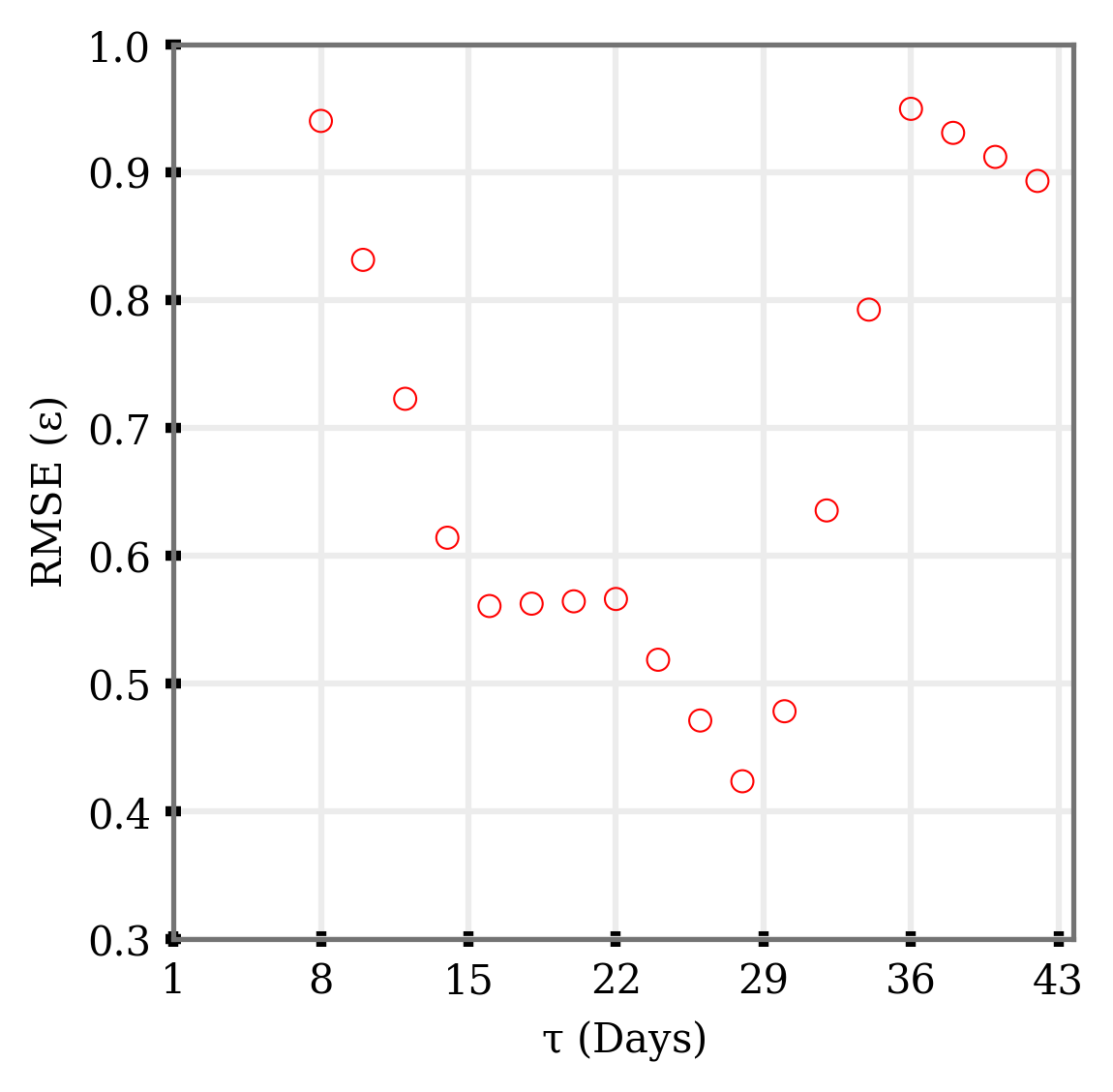}
 \caption{Variation of root mean square error (RMSE, $\epsilon$) in with the length of the training window ($\tau$) in days. Each point of this curve is generated using the average RMSE of two months of data.}
 \label{figure6}
  \end{figure}

The quiet-time and storm-time regressors each use different training windows, the lengths of which were selected to minimize the training error using the mean squared error (MSE) in $\text{K}_\text{p}$ as the loss function.
Figure~\ref{figure6} shows one example of how the MSE varies with the training window (in days) for predictions over two months during 1995.
It can be clearly seen that a training window of $\approx$27 days is optimal at this time, as this captures recurrent structure in the solar wind such as corotating interaction regions \citep{Richardson2012}.
While the deep architecture helps to capture the nonlinear trends in data to provide better accuracy, the Gaussian process mappings are used to provide the error distribution with a mean predicted $\text{K}_\text{p}$. The two dGPR regressors are different in terms of the length of the training window used for forecasting.
The dGPR module dedicated for non-storm, or quiet, conditions has a 27-day training window, whereas the dGPR module for storm conditions uses a 14-day training window.

One of the difficulties in predicting the ``events'' -- i.e., the storm-time $\text{K}_\text{p}$ values -- is that these are typically driven by solar wind transients , which include interplanetary CMEs and corotating interaction regions (CIRs) \citep[see, e.g.,][]{kilpua_small_2009,zhang18transients}, with the largest storms driven by CMEs \citep{borovsky06}. The in-situ solar wind measurements from an L1 monitor do not convey the information required to predict the occurrence of these transients for a 3-hour-ahead prediction of $\text{K}_\text{p}$, or for longer prediction horizons. For this reason, we perform a preliminary investigation in which we include information that may encode the likelihood of CME eruption. Following \citet{winter_estimate_2014-1} we use X-ray flux data from the NOAA GOES platforms as a measure of possible solar activity.

To test whether the inclusion of a proxy for solar activity improves our ability to predict storm--time Kp, we constructed two different prediction systems. The first was trained only on OMNI solar wind parameters ($\mu^\text{OMNI}$). The second added the X-ray background (B) and the flux ratio (R) as extra model parameters ($\mu^{\text{OMNI}^\text{+}}$).
When using the X-ray data we add B and R as model features to the LSTM classifier as well as the storm-time regressor. Note that we do not use the X-ray data for the quiet-time regressor.
Both the models are validated and evaluated against 10 years of $\text{K}_\text{p}$ (2001-2010), in addition to a specific storm-time validation using 38 intervals listed in Table 3.

\begin{center}
\begin{longtable}{lrlrc}
\caption{List of Storm Events}\\
\toprule
\textbf{Start date} & \textbf{Start time} & \textbf{End date} & \textbf{End time} & \textbf{Max. $\text{K}_\text{p}$} \\
\midrule
\endfirsthead
\multicolumn{4}{c}%
{\tablename\ \thetable\ -- \textit{Continued from the previous page}} \\
\toprule
\textbf{Start date} & \textbf{Start time} & \textbf{End date} & \textbf{End time} & \textbf{Max. $\text{K}_\text{p}$} \\
\midrule
\endhead
\midrule \multicolumn{4}{r}{\textit{Continued on next page}} \\
\endfoot
\bottomrule
\endlastfoot
19 March 2001 &	1500 &	21 March 2001 &	2300 &	$7^\text{+}$ \\ 
31 March 2001 &	400 &	1 April 2001 &	2100 &	$9^\text{-}$ \\ 
18 April 2001 &	100 &	18 April 2001 &	1300 &	$7^\text{+}$ \\ 
22 April 2001 &	200 &	23 April 2001 &	1500 &	$6^\text{+}$\\ 
17 August 2001 &	1600 &	18 August 2001 &	1600 &	$7$ \\ 
30 September 2001 &	2300 &	2 October 2001 &	0 &	 $6^\text{-}$\\ 
21 October 2001 &	1700 &	24 October 2001 &	1100 &	$8^\text{-}$ \\ 
28 October 2001 &	300 &	29 October 2001 &	2200 &	$7^\text{-}$ \\ 
23 March 2002 &	1400 &	25 March 2002 &	500 &	$6$ \\ 
17 April 2002 &	1100 &	19 April 2002 &	200 &	$7^\text{+}$ \\ 
19 April 2002 &	900 &	21 April 2002 &	600 &	$7^\text{-}$ \\ 
11 May 2002 &	1000 &	12 May 2002 &	1600 &	$7^\text{-}$ \\ 
23 May 2002 &	1200 &	24 May 2002 &	2300 &	$8^\text{+}$ \\ 
1 August 2002 &	2300 &	2 August 2002 &	900 &	$5^\text{+}$ \\ 
4 September 2002 &	100 &	5 September 2002 &	0 &	 $6^\text{+}$\\ 
7 September 2002 &	1400 &	8 September 2002 &	2000 &	$7^\text{+}$ \\ 
1 October 2002 & 600 &	3 October 2002 	&800 &	$7^\text{+}$ \\ 
20 November 2002 &	1600 &	22 November 2002 &	600 	& $6$\\ 
29 May 2003 &	2000 &	30 May 2003 &	1000 &	$5^\text{+}$ \\ 
17 June 2003 &	1900 &	19 June 2003 &	300 &	$6$ \\ 
11 July 2003 &	1500 &	12 July 2003 &	1600 &	$6$ \\ 
17 August 2003 & 	1800 &	19 August 2003 &	1100 &	$7^\text{+}$ \\ 
20 November 2003 & 	1200 &	22 November 2003 &	0 &	 $9^\text{-}$\\ 
22 January 2004 &	300 &	24 January 2004 &	0 &	$7$ \\ 
11 February 2004 &	1000 &	12 February 2004 &	0 &	 $6^\text{+}$\\ 
3 April 2004 &	1400 &	4 April 2004 &	800 &	$6^\text{+}$ \\ 
22 July 2004 &	2000 &	23 July 2004 &	2000 &	 $7$\\ 
24 July 2004 &	2100 &	26 July 2004 &	1700 &	$6$ \\ 
26 July 2004 &	2200 &	30 July 2004 &	500 &	$7^\text{+}$ \\ 
30 August 2004 &	500 &	31 August 2004 &	2100 & $7$	 \\ 
11 November 2004 &	2200 &	13 November 2004 &	1300 &	$5^\text{+}$ \\ 
21 January 2005 &	1800 &	23 January 2005 &	500 &	$8$ \\ 
7 May 2005 &	2000 &	9 May 2005 &	1000 &	 $8^\text{+}$\\ 
29 May 2005 &	2200 &	31 May 2005 &	800 &	 $8^\text{-}$\\ 
12 June 2005 &	1700 &	13 June 2005 &	1900 & $7^\text{+}$\\ 
31 August 2005 &	1200 &	1 September 2005 &	1200 & $7$\\ 
13 April 2006 &	2000 &	14 April 2006 &	2300 &	$7$\\ 
14 December 2006 &	2100 &	16 December 2006 &	300 &	$8^\text{-}$\\ 
\end{longtable}
\label{tab:3}
\end{center}

\section{Results}
In this section, we present model forecasts and quantitative comparison of predicted $\text{K}_\text{p}$, comparing the models with and without the GOES X-ray data. We further describe a simple method to exploit the uncertainty bounds of the predicted $\text{K}_\text{p}$ to provide a probabilistic geomagnetic storm prediction. Finally, we analyze the performance of the probabilistic $\text{K}_\text{p}$ prediction models.

\subsection{$\text{K}_\text{p}$ Forecast Models}\label{sec:forecastmodels}
\begin{figure}[h]
 \centering
 \includegraphics[scale=0.75]{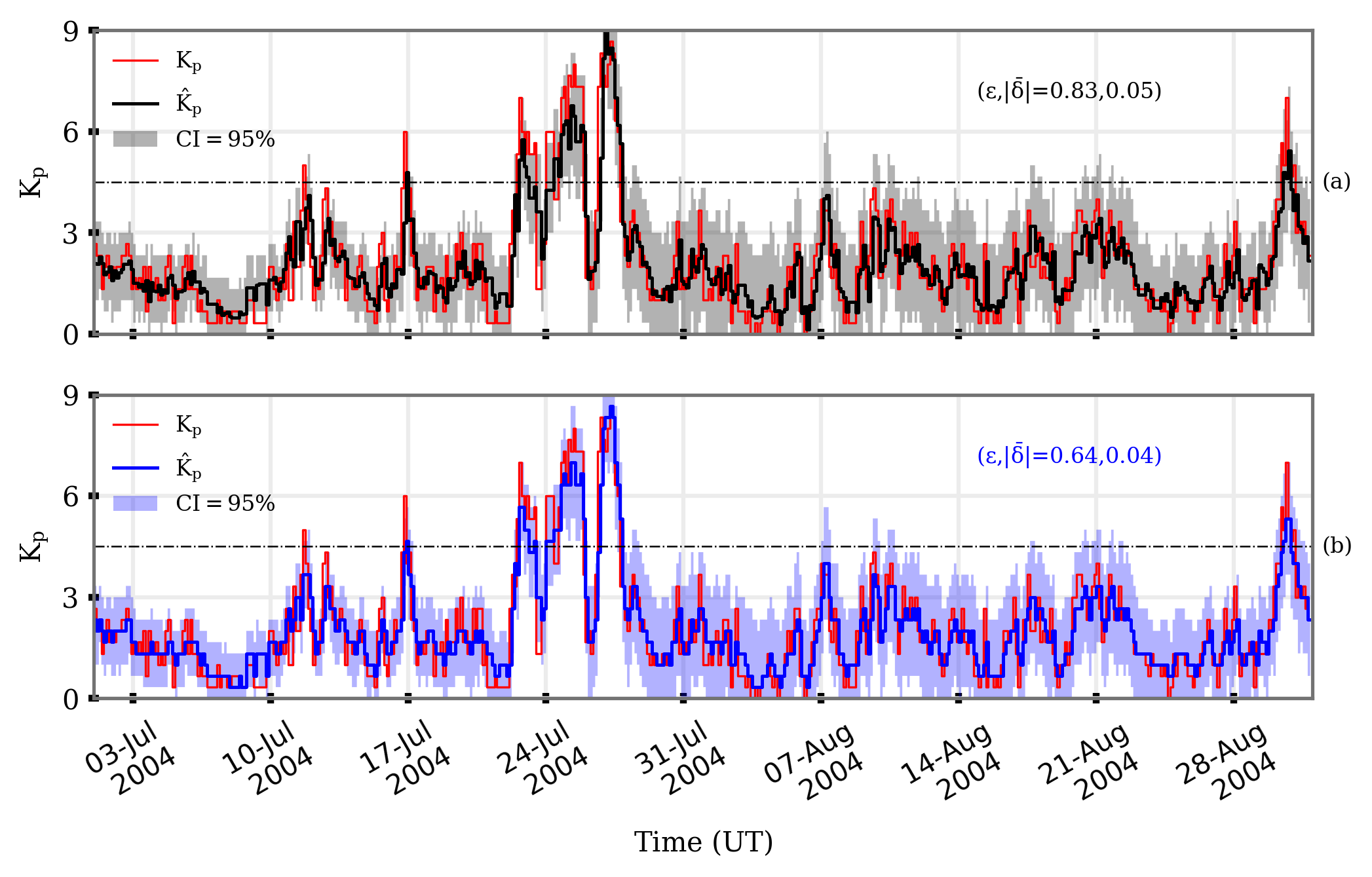}
 \caption{3$-$hour forecast of $\text{K}_\text{p}$ using LSTM classifier \& Deep Gaussian Process Regression (Deep GPR) for a solar maximum period ($\text{1}^\text{st}$ July--$\text{31}^\text{st}$ August, 2004). Panels: (a) prediction from the model using OMNI solar wind data and (b) prediction from the model using OMNI solar wind data and GOES solar flux data. Blue and black dots in panels (a) and (b) are mean predictions and red dots show observed $\text{K}_\text{p}$, respectively. Light blue and black shaded regions in panels (a) and (b) respectively show 95\% confidence interval. RMSE ($\epsilon$) and MAE ($|\bar{\delta}|$) are mentioned inside each panel.}
 \label{figure7}
  \end{figure}

As the first step, we present 3-hour ahead predicted $\text{K}_{p}$ during two months of 2004. Panels (a) and (b) of Figure~\ref{figure7} shows predicted $\text{K}_\text{p}$ with a 95\% confidence interval using models $\mu^\text{OMNI}$ and $\mu^\text{OMNI+}$, respectively. The horizontal dashed black line shows the storm versus non-storm demarcation line.
The root mean square error ($\epsilon$) and mean absolute error ($|\bar{\delta}|$) for this 2 month interval are given each panel as annotations. In the figure, we have discretized the $\text{K}_\text{p}$ values and the 95\% confidence interval bounds by rounding them to the nearest valid $\text{K}_\text{p}$ values~\citep[see section 4.2 of][]{morley18swmf}.
We have chosen this time interval as an example to showcase the ability of the model to capture the dynamics of both storm-time and quiet-time $\text{K}_\text{p}$.
Examining Figure~\ref{figure7}, it is visually apparent that both the models can capture the change in $\text{K}_\text{p}$ during the transitions into, and out of, storm-time. The error metrics given in each panel suggest that models $\mu^\text{OMNI}$ and $\mu^\text{OMNI+}$ are comparable in their performance. However, a more detailed analysis is required to allow us to draw firm conclusions and assess the impact of including the X-ray flux data. Specifically, as the intent of including the X-ray flux is to better capture storm-time transients, we need validation methods that allow us to determine the performance as a function of activity level. 

We have conducted a thorough head-to-head test of two $\text{K}_\text{p}$ forecast models, $\mu^\text{OMNI}$ and $\mu^\text{OMNI+}$, using predictions across our validation data set (from January 2001 through December 2010).
We also compare the model predictions for 38 storm-time intervals (listed in Table 3). Summary statistics for the different models are presented in Table 4.
When comparing the models across the full validation set, the error metrics are nearly identical between the model variants. The RMSE, MAE and correlation coefficient for both of our models show similar performance to the models of \citet{boberg_real_2000} and \citet{bala_improvements_2012}. On the full data set our model does not perform as well as that of \citet{tan_geomagnetic_2018}.
However, in addition to generating a probabilistic forecast, we seek to answer the question of whether including GOES X-ray data provide a meaningful improvement in the prediction of storm-time $\text{K}_\text{p}$ intervals.
Looking at the same metrics for the 38 storm-time intervals, a different picture emerges.
The model variant incorporating X-ray flux data ($\mu^\text{OMNI+}$) outperforms the standard model by a substantial margin.
The RMSE of $\mu^\text{OMNI+}$ is 0.9 and the MAE is 0.67, showing that typical storm-time Kp predictions are within 1 Kp interval of observation.
Of particular importance is that the correlation coefficient increases for $\mu^\text{OMNI+}$ relative to the performance across the full validation set. Here we note that the correlation coefficient can be considered a measure of potential performance, as it neglects conditional and unconditional bias \citep{Murphy1995}.

\begin{figure}[h]
 \centering
 \includegraphics[scale=0.9]{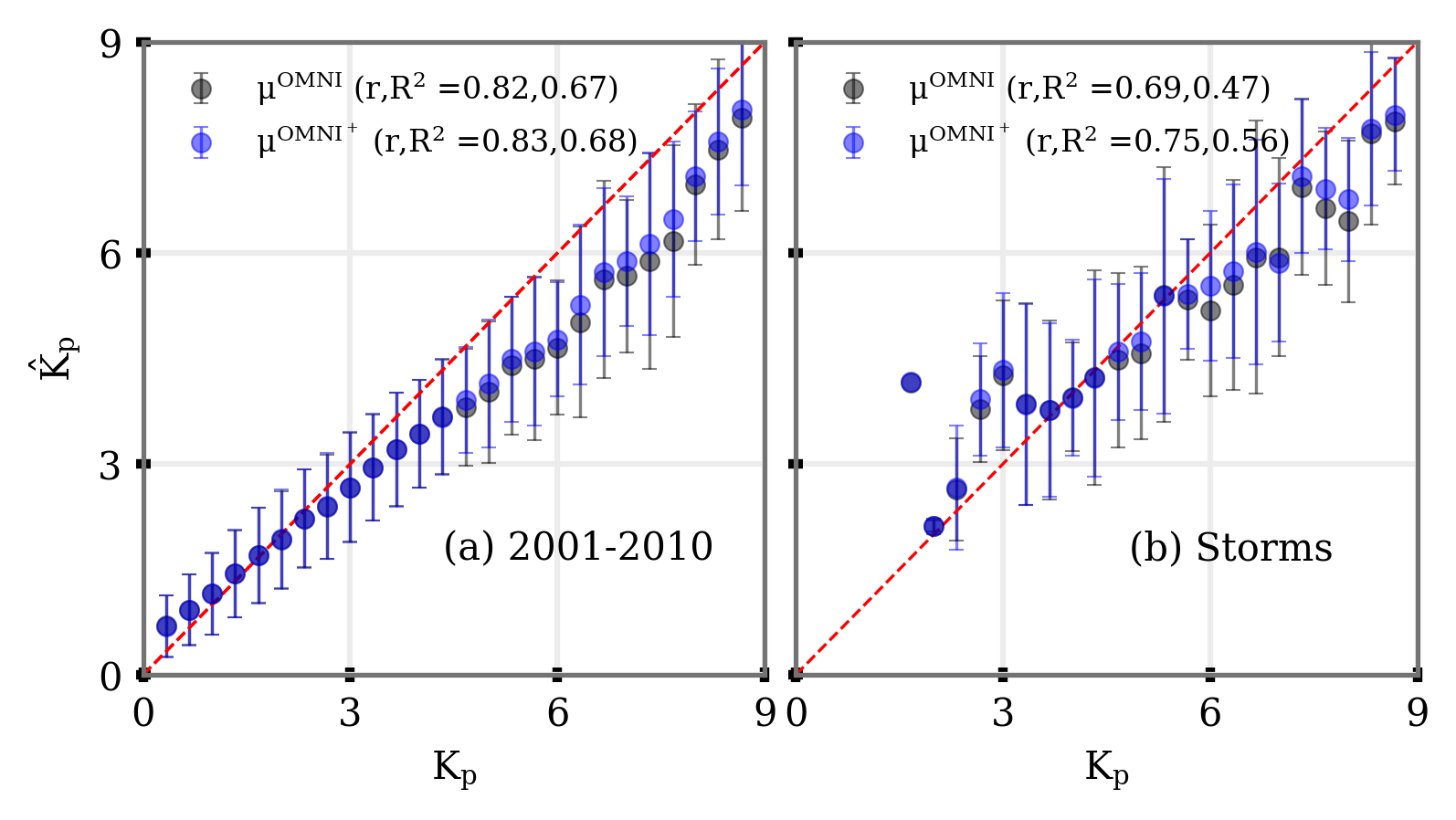}
 \caption{The performance comparison of $\mu^\text{OMNI}$ and $\mu^\text{OMNI+}$ models which predict $\text{K}_\text{p}$ 3-hour ahead. Panels present performance of $\mu^\text{OMNI}$ (in black) and $\mu^\text{OMNI+}$ (in blue) models for (a) 10 years of prediction and (b) 38 storm--time prediction (listed in the Table~3). In each panel official (Postdam) $\text{K}_\text{p}$ is plotted on the x-axis and the model prediction is plotted on the y-axis. Perfect predictions would lie on the line with a slope of one. The error bar indicates one standard deviation and the correlation coefficient and coefficient of determination are mentioned inside each panel.}
 \label{figure8}
  \end{figure}

To graphically display the model bias across these two validation sets, Figure~\ref{figure8} plots observed $\text{K}_\text{p}$ against predicted $\text{K}_\text{p}$. Panel (a) shows the comparison across the full 10-year validation set and panel (b) shows the comparison for the 38 storm intervals.
Black and blue colors represent predicted $\text{K}_\text{p}$ using $\mu^\text{OMNI}$ and $\mu^\text{OMNI+}$, respectively. The circles give the mean predicted $\text{K}_\text{p}$ and the vertical lines represent 1-$\sigma$ error bars associated with that predicted $\text{K}_\text{p}$ level.
As above, the predictions from both models are comparable when we use the full validation set (Figure~\ref{figure8}(a)) and do not account for the activity level.
For $\text{K}_\text{p} \leq 3^{+}$ the predictions show little to no bias; the mean predicted $\text{K}_\text{p}$ is nearly identical to the observed value during quiet times.
At higher levels of geomagnetic activity, we see a clear tendency for the mean predicted $\text{K}_\text{p}$ to be lower than the observation. That is, high values of $\text{K}_\text{p}$ tend to be underpredicted by both models.
Comparing the two models shows a slight improvement in the bias at higher $\text{K}_\text{p}$ values using the $\mu^\text{OMNI+}$ model, but the most visible improvement using this display format is in the smaller error bars on the $\mu^\text{OMNI+}$ predictions.

Table~4 presents a summary of overall fit statistics, both both model, for the 2001-2010 data set as well as the storm-time data set. On both data sets there is an improvement in the fit when adding the X-ray flux data to the model. Because the correlations have the observations in common, we test whether the improvement is significant using a test for correlated correlation coefficients \citep{steiger80,revelle20}, where we use the effective number of samples ($n_{eff}$ where n$_{eff} < n$) estimated by correcting for the lag-1 autocorrelation \citep[see, e.g.,][]{wilks06}. The improvements in $r$ for both models are statistically significant, with a p-value of $4.8\mathrm{x}10^{-5}$ for the 2001-2010 data set and $p < 10^{-5}$ for the storm-time data set.
Given the results presented in both Table~\ref{tab:4} and Figure~\ref{figure8}, we conclude that including X-ray flux information provides information that improves $\text{K}_\text{p}$ prediction accuracy during geomagnetically active intervals.

\begin{table}[!htb]
    \centering
    \begin{tabular}{cccccc}
        \toprule
         \textbf{Forecast Model} & \textbf{Case} & \textbf{r} & R\textbf{MSE} ($\epsilon$) & \textbf{MAE} ($|\bar{\delta}|$) & \textbf{$\text{R}^\text{2}$}\\ \midrule
         $\mu^\text{OMNI}$ & 2001$-$2010 & 0.82 & 0.78 & 0.59 & 0.67 \\ 
         $\mu^\text{OMNI+}$ & 2001$-$2010 & 0.83 & 0.77 & 0.58 & 0.68 \\
         $\mu^\text{OMNI}$ & Storms & 0.69 & 1.48 & 1.11 & 0.29 \\ 
         $\mu^\text{OMNI+}$ & Storms & 0.75 & 0.9 & 0.67 & 0.56 \\
         \bottomrule
    \end{tabular}
    \caption{Table Showing the Prediction Summary of the $\mu^\text{OMNI}$ and $\mu^\text{OMNI+}$ models during 2001--2010 and geomagnetic storms listed in Table 3.}
    \label{tab:4}
\end{table}\textbf{}

However, this analysis only uses the mean prediction and neglects the fact that we have used a probabilistic regressor.
So, how can we use the uncertainties provided by the dGPR prediction and how well do our probabilistic predictions perform?


\subsection{Probabilistic Storm Forecast}
\begin{figure}[h]
 \centering
 \includegraphics[scale=0.75]{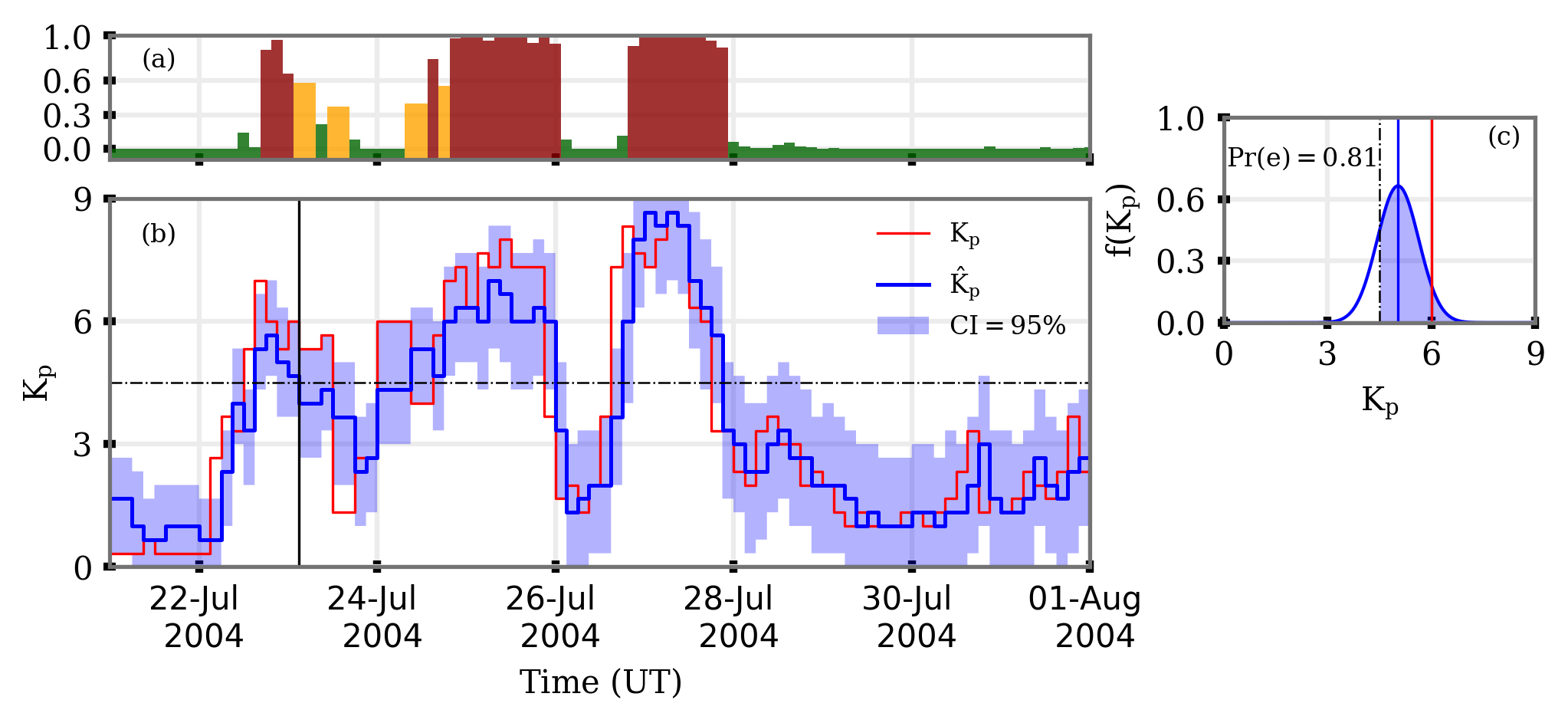}
 \caption{3-hour forecast (using $\mu^\text{OMNI+}$ model) showing (a) probability of geomagnetic storms, (b) $\text{K}_\text{p}$ ($\text{22}^\text{nd}$ July--$\text{31}^\text{st}$ July, 2004) and (c) an illustration of the method to extract probability of storm occurrence for one prediction marked by vertical black line in panel (b). Black dashed lines in panels (b) and (c) represent the threshold $\text{K}_\text{p}$=$5^-$, red and blue thin lines in panel (c) are observed $\text{K}_\text{p}$, and predicted mean respectively. Panel (b) is in the same format with Figure~\ref{figure7}. The shaded region in panel (c) provides probability of geomagnetic storm (Pr[e]=0.81, for details refer text).}
 \label{figure9}
  \end{figure}

Here we describe how we exploit the uncertainty in predicted $\text{K}_\text{p}$ to provide a probabilistic geomagnetic storm forecast using the SW-driven model $\mu^\text{OMNI}$ as an example.
Figure~\ref{figure9} illustrates the probabilistic prediction of both the $\text{K}_\text{p}$ index and of geomagnetic storm occurrence. Figure~\ref{figure9}(a) shows a ``traffic light'' display which gives the probability of $\text{K}_\text{p}$ exceeding 5$^{-}$, which we use to delineate storm-time, following the NOAA G-scale of geomagnetic storms (refer Table~\ref{tab:1}).
The color represents the likelihood of storm activity: green is Pr $\leq 0.3$, yellow is $0.33<\mathrm{Pr}\leq0.66$, and red marks intervals where the probability of geomagnetic storm conditions exceeds 0.66. Note that, Pr is the probability of geomagnetic storm, using the NOAA SWPC definition of $\text{K}_\text{p}\geq\text{5}^\text{-}$ as the threshold for geomagnetic storm.

Figure~\ref{figure9}(b) presents 10 days ($\text{21}^\text{st}$--$\text{31}^\text{st}$ July 2004) of 3-hour ahead predictions of $\text{K}_\text{p}$, using $\mu^\text{OMNI+}$ model (cf. Figure 7(b)). The horizontal dashed line is at $\text{K}_\text{p} = 5^-$ and the vertical bar marks the time of the prediction shown in Figure~\ref{figure9}(c).
Figure~\ref{figure9}(c) illustrates how to calculate the probability of geomagnetic storm from the predicted distribution of $\text{K}_\text{p}$. The blue curve gives the output Gaussian probability density function from the dGPR regressor while the blue and red vertical lines represent the mean prediction and the observed $\text{K}_\text{p}$.
The vertical dashed black line marks $\text{K}_\text{p}$=$5^-$ and the integral of the shaded area is the probability of exceeding the threshold. 

In a non-probabilistic model, we would simply have a binary outcome of storm or non-storm. Following this method, we see that the prediction of a probability distribution gives us both uncertainty bounds on our prediction of the actual $\text{K}_\text{p}$ as well as the probability of exceeding a given threshold in $\text{K}_\text{p}$.

To assess the probabilistic storm prediction (i.e., the probability of exceeding a threshold), we will examine binary event forecasts.
For this we convert each probabilistic prediction into a prediction of ``event'' or ``no event''. For this, we need to choose a probability threshold.
As our predicted probability distribution is Gaussian, the mean prediction is also the $\text{50}^\text{th}$ percentile. Simply ignoring the predicted distribution and using the mean value is equivalent to using a threshold of 0.5.
We can similarly convert the observed $\text{K}_\text{p}$ to a series of events and non-events, and can subsequently determine whether the prediction was a true positive, false positive, false negative or true negative (see section 2.2). 

\begin{figure}[h]
 \centering
 \includegraphics[scale=0.75]{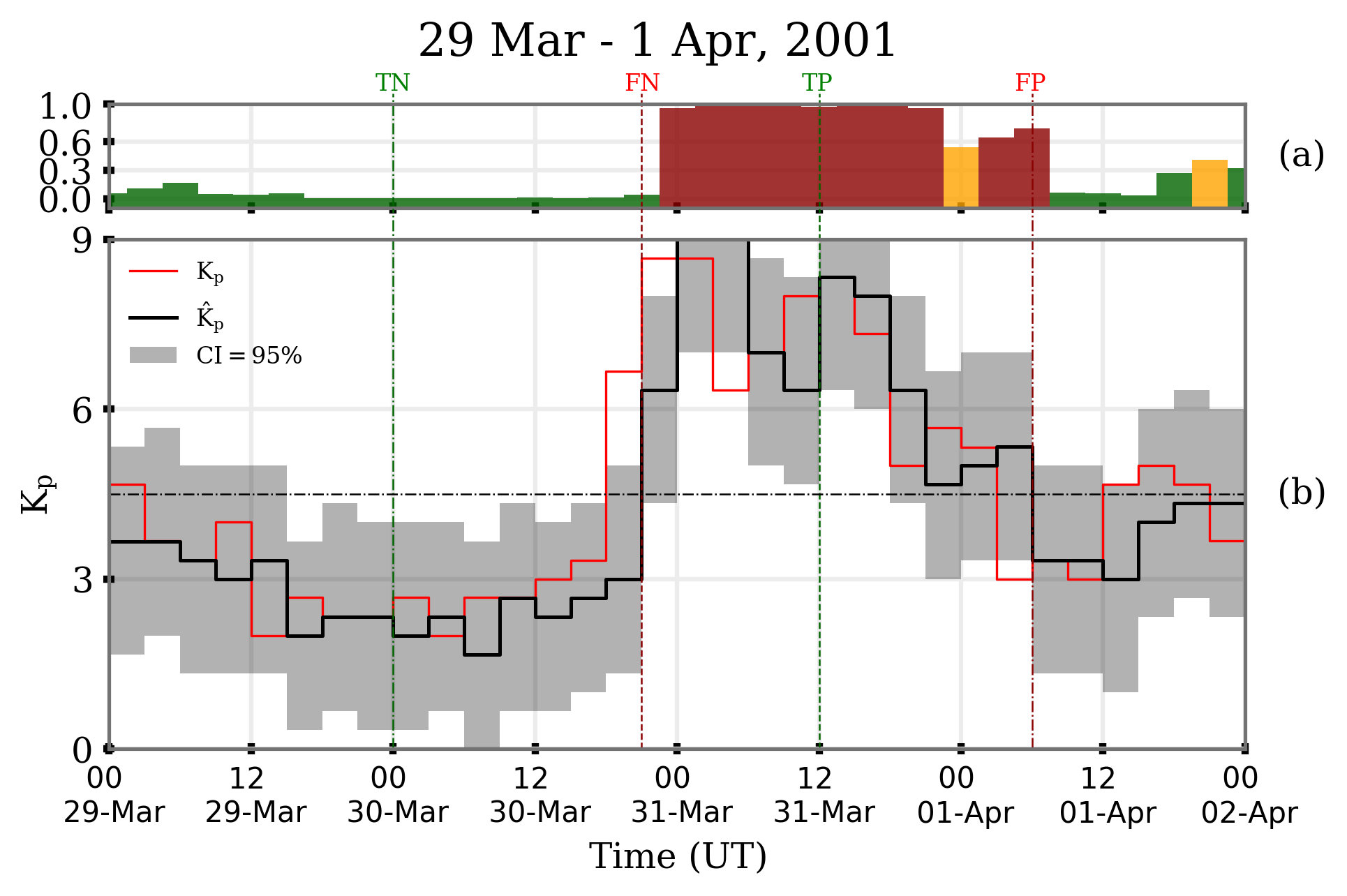}
 \caption{Example model predictions using $\mu^\text{OMNI}$ model showing True Positive (TP), False Positive (FP), False Negative (FN), and True Negative(TN) predictions, mentioned by vertical lines. Top and bottom panels show the probability of geomagnetic storms and $\text{K}_\text{p}$  with uncertainty bounds (shaded) region.}
 \label{figure10}
  \end{figure}

Figure~\ref{figure10} shows $\text{K}_\text{p}$ predictions using the $\mu^\text{OMNI}$ model during a geomagnetic storm at the end of March 2001.
One forecast from each of the true positive (TP), false positive (FP), false negative (FN) and (true negative) TN categories are shown by the vertical lines. Figure~\ref{figure10} is in a similar format to Figure~\ref{figure9}, except that panel (c) is omitted. We use the simpler model for this graphic to illustrate the main effect that our $\mu^{\text{OMNI}^{+}}$ model aims to combat. Specifically, the FP and FN cases are occurring in a specific pattern. At the start of the storm, we see a false negative and at the end of the storm, we see a false positive. This is typical for solar wind-driven models for predictions that are further ahead than the lead time given by the solar wind. In this case, we have a 3-hour lead time to our prediction and so the model has no information to capture the sudden onset of the geomagnetic activity. By the next prediction, the model has ``caught up'' and now correctly predicts a very high likelihood of storm-time.
The inverse is seen, though perhaps less clearly, at the trailing edge of the storm. The model is unable to predict the cessation of storm-level geomagnetic activity, although we do note that the uncertainty bounds include a non-negligible likelihood of $\text{K}_\text{p} < 5^-$.

The model prediction is, therefore, lagging the observation. While this figure shows one example where predicted $\text{K}_\text{p}$ is lagging the observations by 3 hours, most of the storm-time predictions are lagging the observations by at least one 3-hour prediction window. This implies that the model $\mu^\text{OMNI}$ has insufficient information to capture the imminent arrival of a solar wind transient from the L1 data alone, and the prediction is likely to be strongly persistent (giving high weight to the previous value of $\text{K}_\text{p}$) in the model.
By including the X-ray data in the $\mu^{\text{OMNI}^{+}}$ model as a proxy for the likelihood of CME occurrence, we aim to improve storm-time predictions and hopefully combat this ``lag'' effect.

\subsection{Comparison of Probabilistic Predictions}\label{sec:comparisonprobabilistic}
In the previous sections, we described a method to use the uncertainty bound to a predict probabilistic storm forecast. Here we are going to compare the predictions using models ($\mu^\text{OMNI}$ and $\mu^\text{OMNI+}$), using the different metrics defined in section 2.
We begin with the Receiver Operating Characteristic (ROC) curve.
The ROC curve is calculated from the probability of detection (PoD) and the probability of false detection (PoFD) over a range of decision thresholds.
If we make a decision threshold of Pr $=0$, then all predictions lie above it and thus every time is predicted as an event and the PoD and PoFD are both 1.
Conversely, taking a decision threshold of Pr $=1$ leads to no events being predicted, thus PoD and PoFD are both zero. 

\begin{figure}[h]
 \centering
 \includegraphics[scale=1.0]{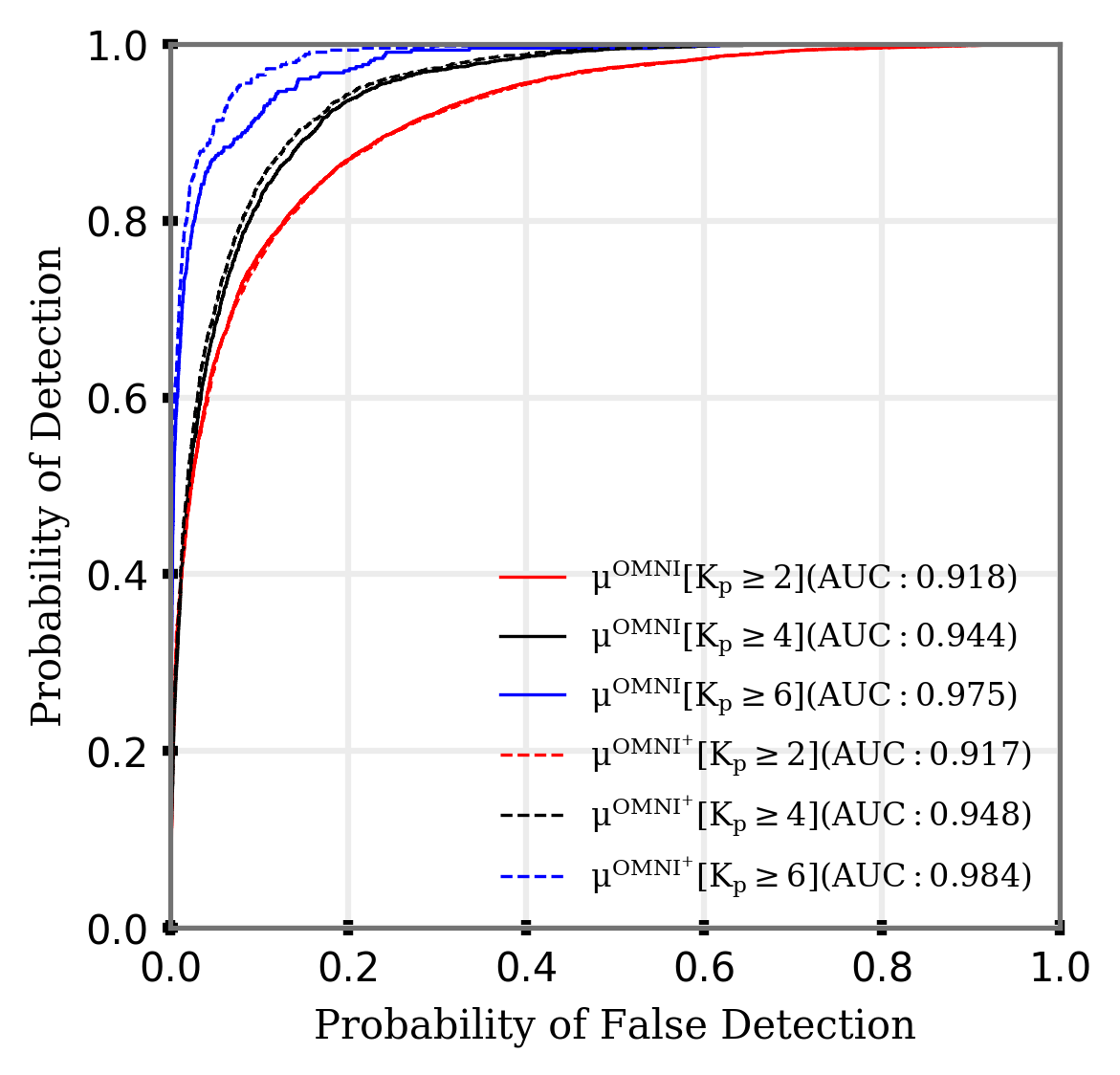}
 \caption{Receiver operating characteristic (ROC) curves showing the relative performance of individual the storm forecasting model $\mu^\text{OMNI}$ (represented by solid curves) and $\mu^\text{OMNI+}$ (represented by dashed curves) with different storm intensity levels (for $\text{K}_\text{p}\geq$ 2, 4, and 6).}
 \label{figure11}
  \end{figure}

Figure~\ref{figure11} presents the ROC curves calculated for both the $\mu^\text{OMNI}$ and $\mu^\text{OMNI+}$ models, using different $\text{K}_\text{p}$ threshold values. The solid lines are the ROC curves from model $\mu^\text{OMNI}$ and the dashed lines are the ROC curves from model $\mu^\text{OMNI+}$. Thresholds of $\text{K}_\text{p}$ = 2, 4 and 6 are shown in red, black and blue, respectively.
We also use the area under the ROC curve (abbreviated as AUC) as a summary measure to compare the performance of our models \citep[cf.][]{Bradley1997}, and the AUC for each ROC curve is given in the figure legend.
For the lower $\text{K}_\text{p}$ threshold values are shown ($\text{K}_\text{p}=$2 or 4) the curves are similar and the AUC are correspondingly similar.
For the higher $\text{K}_\text{p}$ threshold value ($\text{K}_\text{p}=$6) the ROC curves visibly diverge across a broad range of decision thresholds. The AUC for $\mu^\text{OMNI+}$ is higher than that for $\mu^\text{OMNI}$.
This provides qualitative support for the hypothesis that the inclusion of GOES X-ray data has improved the performance of our $\text{K}_\text{p}$ model for high geomagnetic activity.

As the aim of including the X-ray flux data was to potentially provide information relevant to predicting the intervals of higher $\text{K}_\text{p}$ with a longer lead time, we also test the difference between the AUC for $\mu^\text{OMNI+}$ and $\mu^\text{OMNI}$, when $\text{K}_\text{p} \geq 6$. Because the same test data is used for both ROC curves -- the two blue curves in figure~\ref{figure11} -- we use DeLong's nonparametric test for the area under correlated ROC curves \citep{delong88,Sun14} as implemented in the pROC package \citep{robin14}. A two-sided test yielded ($Z=-8.27; p < 2.2\textrm{x}10^{-16}$) showing that the visual difference between the ROC curves for the two models is statistically significant. This confirms the qualitative analysis presented above and supports the hypothesis that including even simple proxies for solar activity can improve the prediction of geomagnetic activity with lead times greater than the L1-to-Earth transit time.

\begin{figure}[h]
 \centering
 \includegraphics[scale=0.75]{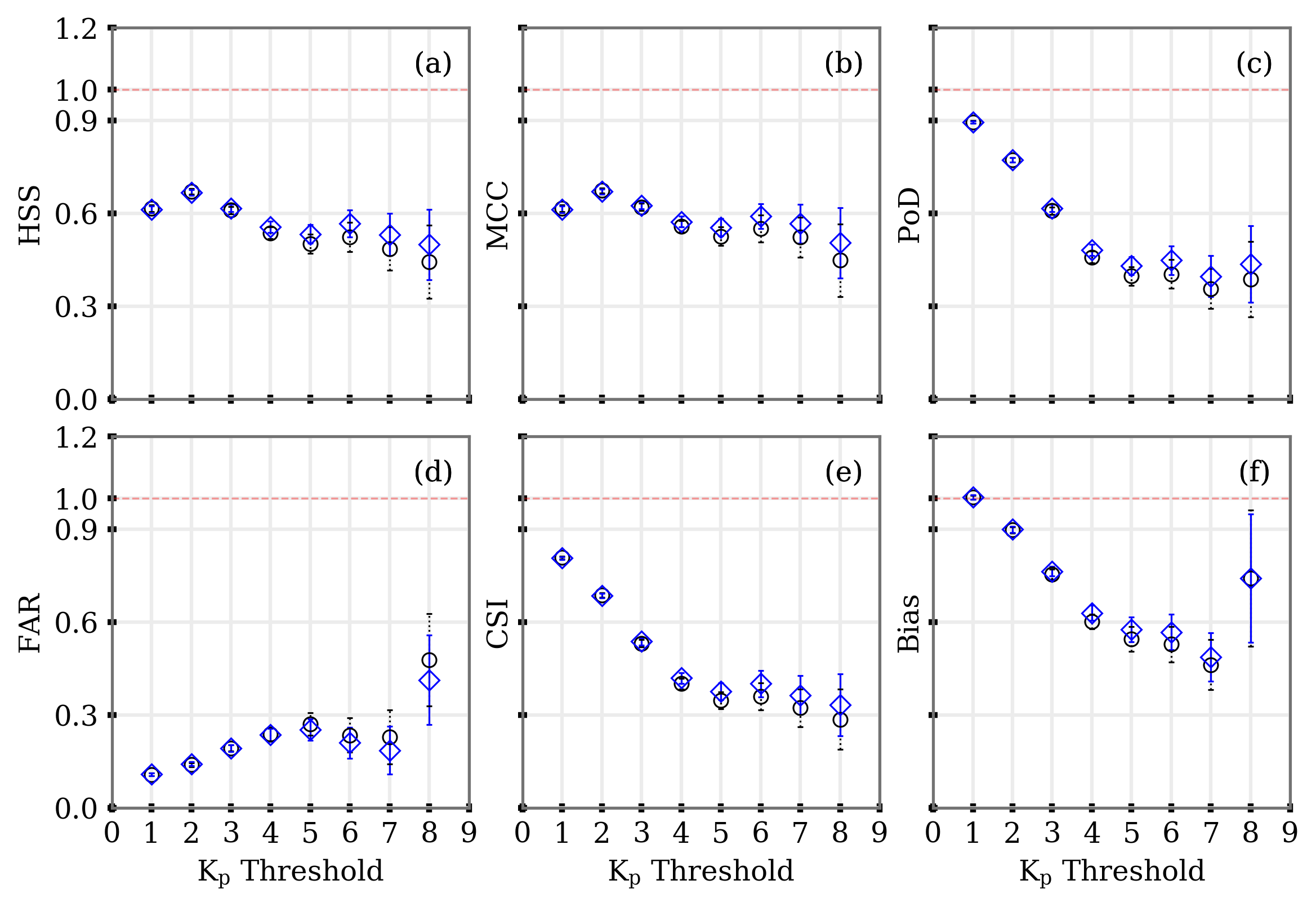}
 \caption{Different performance metrics (a) HSS, (b) MCC, (c) PoD, (d) FAR, (e) CSI, and (f) Bias comparing the two variants of geomagnetic storm forecasting model at different $\text{K}_\text{p}$ thresholds. Model $\mu^\text{OMNI}$ (in black circle) and $\mu^\text{OMNI+}$ (in blue diamonds).}
 \label{figure12}
  \end{figure}

Figure~\ref{figure12} also explores activity-dependent model performance.
Using Pr$=0.5$ as our decision threshold again, we calculate a range of performance metrics (described in section 2.2) while varying the $\text{K}_\text{p}$ threshold used to define an ``event'' \citep[see also][]{Liemohn2018ModelPredictions}.
In each of the panels, the black markers show the results for $\mu^\text{OMNI}$ and the blue markers show the results for $\mu^\text{OMNI+}$.
The error bars show the 95\% confidence interval estimated using 2000 bootstrap resamplings \citep{morley18swmf,PyForecastTools}.
Panel (a) shows the threshold-dependent Heidke Skill Score (HSS), which measures model accuracy relative to an unskilled reference.
Panel (b) shows the Matthews Correlation Coefficient, which can be interpreted in a similar way to a Pearson correlation.
Panel (c) shows the probability of detection (PoD), also called hit rate or true positive rate.
Panel (d) shows the false alarm ratio (FAR), which is the fraction of predicted events that did not occur. The ideal FAR is therefore zero.
Panel (e) shows the critical success index (CSI), which can be interpreted as an accuracy measure after removing true negatives from consideration.
Finally, panel (f) displays the frequency bias, where an unbiased forecast predicts the correct number of events and non-events and scores 1.
As a reminder, the metrics displayed in panels a, b, c and e are positively-oriented, where 1 constitutes a perfect score. FAR (panel d) is negatively-oriented and a perfect model has an FAR of zero. The metrics shown in panels a-e have an upper bound of 1, and this is marked by the red dashed line.
In every measure, the performance between the two models is indistinguishable at low values of $\text{K}_\text{p}$ -- which, as we recall, constitutes the vast majority of geomagnetic conditions -- but as the threshold for identifying an event increases we clearly see improved performance from the $\mu^\text{OMNI+}$ model.
While the confidence intervals substantially overlap for these scores we note that parameter estimates with overlapping confidence intervals can be significantly different \citep{afshartous10}. \added{In other words, while non-overlapping confidence intervals are likely to show that the performance metrics are significantly different, the inverse is not necessarily true. Due to a variety of factors, we cannot assess the significance of the improvement in all performance metrics presented here. Among these are the fact that the metrics are correlated with each other, and we would need to correct for the effect of multiple significance tests. We have instead noted throughout this work where we were able to test for significance and described the consistent improvement in performance metrics.} 
Although the improvement in these metrics is modest, we again conclude that adding the GOES X-ray flux data improves the model's ability to predict geomagnetically active times.

Finally, we assess the reliability of the probability distributions generated by our dGPR models.
In this context, reliability assesses the agreement between the predicted probability and the mean observed frequency. In other words, if the model predicts a 50\% chance of exceeding the storm threshold, is that prediction correct 50\% of the time?

\begin{figure}[h]
 \centering
 \includegraphics[scale=1.0]{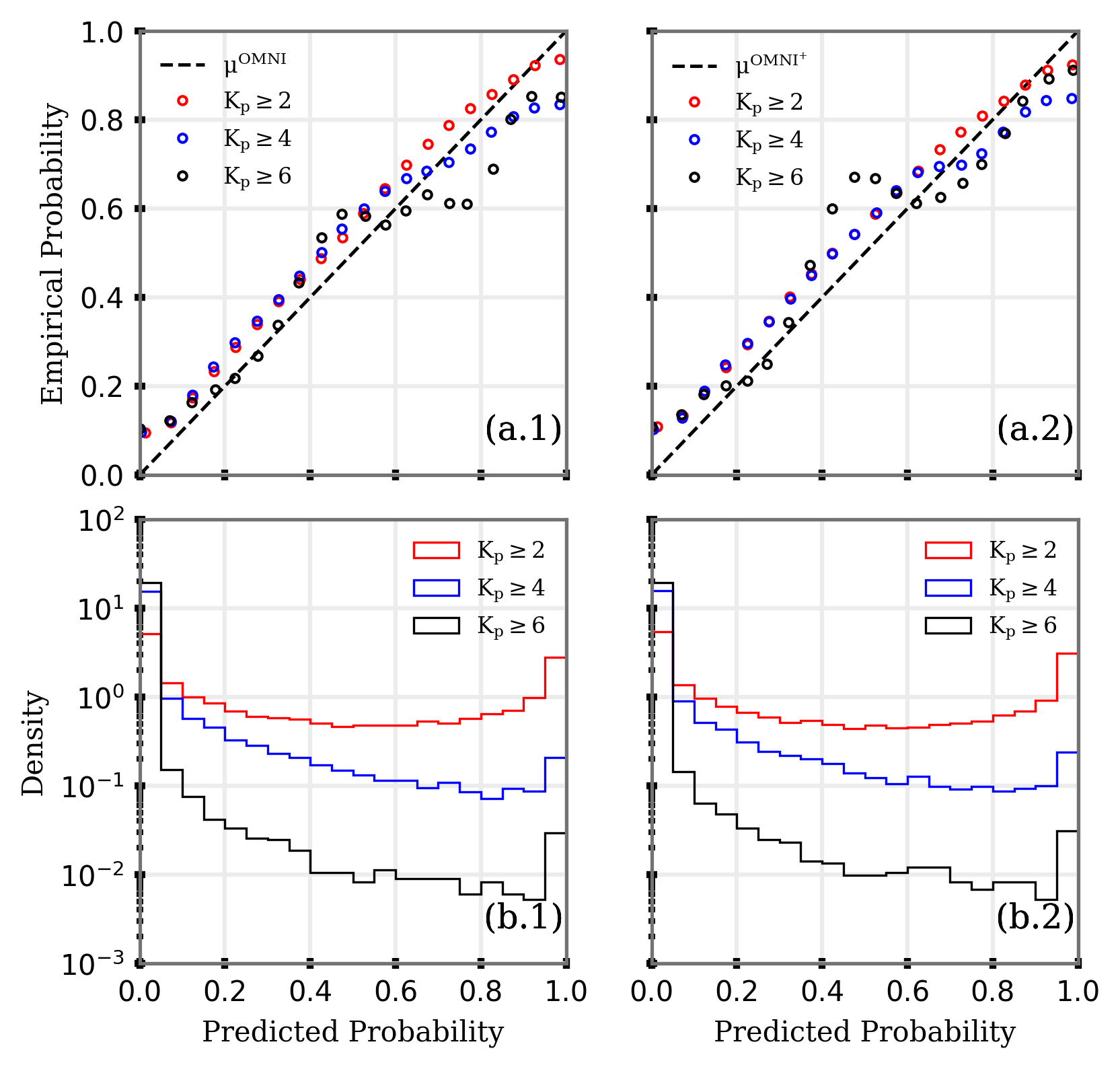}
 \caption{Reliability diagram showing observed frequency of a geomagnetic storm (for $\text{K}_\text{p}\geq$ 2,4, and 6) plotted against the Forecast probability of geomagnetic storms.}
 \label{figure13}
  \end{figure}

Figure~\ref{figure13} presents a reliability diagram of the observed probability of a geomagnetic storm (for different $\text{K}_\text{p}$ threshold levels, i.e. $2,4,6$) plotted against the forecast probability of a geomagnetic storm.
The top row -- panels (a.1) and (a.2) -- presents reliability diagram for models $\mu^{OMNI}$ and $\mu^{OMNI+}$, respectively. In this figure red, blue and black lines represent geomagnetic storm thresholds of $\text{K}_\text{p}$ = 2, 4 and 6 respectively.
A perfectly reliable forecast should lie on the $x=y$ line (black dashed line).
For smaller chances of geomagnetic storms, both forecast models are reliable in their probabilistic predictions.
As the predicted probability increases so do the tendency for the predicted probability to be higher than the observed probability. That is, the model tends to over-forecast slightly.
Comparing the reliability of $\mu^{OMNI}$ to that of $\mu^{OMNI+}$, we see similar results for activity thresholds of $\text{K}_\text{p} = 2$ and 4.
However, the $\mu^{OMNI+}$ model predictions for a storm-time threshold of $\text{K}_\text{p} = 6$ are slightly more reliable than for its simpler counterpart.

The panels in the bottom row of Figure~\ref{figure13} are histograms showing the frequency of forecasts in each probability bin, also known as refinement distributions. These indicate the sharpness of the forecast, or the ability of the forecast to predict extremes in event probability.
For example, a climatological mean forecast would have no sharpness and a deterministic model (i.e., a prediction with a delta function probability distribution) would be perfectly sharp, only ever predicting probabilities of zero or one.
Ideally, a model would have both sharpness and reliability in its predicted probabilities.
The refinement distributions presented here show that both $\mu^{OMNI}$ and $\mu^{OMNI+}$ display sharpness, with local peaks near probabilities of zero and one.

Both models exhibit high sharpness, which can be interpreted as the confidence of the model in its event prediction \citep{wilks06}. Further, both models perform similarly for lower $\text{K}_\text{p}$ thresholds. The $\mu^{OMNI+}$ model, when using an event threshold of $\text{K}_\text{p} \geq 6$, has slightly improved calibration over the $\mu^{OMNI}$. The addition of the solar X-ray flux data has consistently improved performance when assessing its performance in a deterministic setting, and here is shown to improve the calibration of the model at high activity levels without impact to the sharpness of the model.
This analysis further supports the performance of our probabilistic model and confirms that the GOES X-ray data adds value to our $\text{K}_\text{p}$ prediction model.

\section{Discussion}
In this study, we presented a novel, probabilistic, geomagnetic storm forecast model that predicts 3 hours ahead. Our model structure combined an LSTM classifier and dynamically-trained deep Gaussian processes to generate predictions of $\text{K}_\text{p}$ with an associated probability distribution.
To test whether a simple, operationally-available data set could improve predictions of geomagnetic storm times, we trained two variants of our model: the first used only solar wind data and historical values of $\text{K}_\text{p}$; the second added X-ray fluxes from the NOAA GOES satellites, as a proxy for magnetic complexity at the Sun.
Using a variety of model validation methods, we have confirmed that including the X-ray data enhances the performance of the forecast model at times of high geomagnetic activity. Due to the low number of samples (at high $\text{K}_\text{p}$ levels) for model testing, many measures of model performance suggest an improvement in the model performance at high activity levels but statistical significance could not be demonstrated. Significance tests of the improvement in the correlation coefficients and the change of the ROC AUC show that there is a quantified, statistically-significant improvement in the model performance when GOES X-ray flux data is included. In this section, we further analyze the performance metrics and compare them with prior studies.

Although exact comparisons should not be made as we use different data sets for model validation, we place our results in the context of previous work.
In comparison with some earlier models \citep[e.g.][]{boberg_real_2000,wing_kp_2005,bala_improvements_2012} our models typically perform well, with an RMSE of $0.77$. 
The performance, as measured by RMSE, is not as good as the RMSE for the 3\,hr-ahead predictions of \citet{Zhelavskaya2019} (RMSE = $0.67$) and \citet{tan_geomagnetic_2018} (RMSE = $0.64$).
To assess the performance of their model when predicting geomagnetic storm intervals (defined as $\text{K}_\text{p} \geq 5$), \citet{tan_geomagnetic_2018} has calculated the F1-score.
This binary event metric is the harmonic mean of the \textit{precision} and \textit{recall}, using the nomenclature of machine learning literature.
Using terminology from statistical literature, the precision is perhaps better known as the positive predictive value and represents the fraction of predicted positives that were correct.
Similarly, the recall is the probability of detection and represents the fraction of observed events that were correctly predicted.
The F1-score for the \citet{tan_geomagnetic_2018} model was $0.55$.
Our initial model ($\mu^{OMNI}$) gave an F1-score of $0.56$, while our model including the solar X-ray flux data ($\mu^{OMNI+}$) gave an F1-score of $0.6$.

Recent studies mainly focused on the predictive skill of the $\text{K}_\text{p}$ forecast models, whereas, in this paper, we aim to provide a probabilistic prediction of $\text{K}_\text{p}$ without compromising the predictive skill of the model.
We have further demonstrated that including a simple, operationally-available proxy for the likelihood of solar activity improves the prediction of geomagnetic storms.
The inability of $\text{K}_\text{p}$ prediction models to predict larger storms ($\text{K}_\text{p} \geq 5$) well from L1 solar wind data has previously been discussed in the literature \citep[see, e.g.,][]{Zhelavskaya2019}, and this work shows that including solar X-ray flux can directly improve the prediction of high levels of geomagnetic activity.
In this work we found that including solar X-ray flux in our model features reduces the overall RMSE by 0.01, from 0.78 to 0.77. At the same time the correlation coefficient increased by a small but statistically significant amount (from 0.82 to 0.83). Importantly, for the storm-time test data the RMSE was reduced by 0.58, from 1.48 to 0.9, and the correlation coefficient increased from 0.69 to 0.75. For details of the results and the significance testing, see Table~\ref{tab:4} and section~\ref{sec:forecastmodels}. Similarly, we note that analyzing the area under the ROC curve shows a significant improvement in the probabilistic predictions of $\text{K}_\text{p}$, for high activity levels, when X-ray flux is included (see section~\ref{sec:comparisonprobabilistic}). These comparisons show that inclusion of solar flux can enhance the storm time forecasting capability without diminishing the performance during less active periods.

Although we present only one sample model architecture, we use this to highlight a straightforward method by which uncertainty bounds can be predicted using machine-learning models, and also improve predictions of intervals of high geomagnetic activity.
This clear demonstration that the X-ray flux data meaningfully improves our prediction of geomagnetic storms strongly suggests that future work including solar data sources are a promising way to extend the meaningful forecast horizon for high levels of geomagnetic activity.
However, other operationally-available data sources exist that are likely to carry more information about magnetic complexity at the Sun (e.g., solar magnetograms \citep{Arge2010}), and hence using these will further improve the prediction of both the CME- and non-CME-driven geomagnetic activity.
Further work is planned to investigate this work as well as incorporating other recent advances that will help improve the fidelity of our predictive model.

We also note that \citet{Zhelavskaya2019} explored methods for selecting model features and reducing a large number of candidate features to a smaller selection of those that are most important.
Relative to their work we use a small number of features, all of which were selected based on a physically-motivated interpretation and subject to the constraint of being products generally available operationally.
Applying their feature selection methods and further developing the architecture, such as using convolution neural network to process and extract CME features from 2-dimensional solar X-ray and magnetogram data, would likely yield immediate improvements in the model performance.

\section{Conclusion}
The two main objectives addressed by this work were to: 1. build a probabilistic $\text{K}_\text{p}$ forecast model; and 2. test whether the inclusion of operationally-available proxies for solar activity could improve the prediction of geomagnetic storms (using $\text{K}_\text{p}$, following the NOAA G-scale).

We presented a two-layer architecture, using an LSTM neural network to predict the likelihood of storm-time and deep Gaussian Process Regression models to predict the value of $\text{K}_\text{p}$ including uncertainty bounds.
We then exploited these uncertainty bounds to provide a probabilistic geomagnetic storm forecast.
Our analysis demonstrates that this architecture can be used to build probabilistic space weather forecast models with good performance.

Further, we tested two variants of our model that differed only in the input parameters (``features''). The first used only upstream solar wind data and the second added solar X-ray flux data from the GOES spacecraft.
Analysis of the predictions and the errors, for both the values of $\text{K}_\text{p}$ as well as the probability of exceeding a threshold in $\text{K}_\text{p}$, showed that the addition of X-ray flux data resulted in significant model performance improvements during geomagnetically active periods.
The model using X-ray flux data had a significantly higher correlation coefficient on the storm-time test data (increased from 0.69 to 0.75), with other performance metrics showing improvements. The RMSE on the storm-time data set decreased from 1.48 to 0.9. This improvement in model performance was also seen across all contingency table-based metrics, with improvements in skill and reductions in false alarms.
Similarly, the probabilistic predictions were shown to be significantly better by testing the difference in the area under the ROC curve. The probabilistic predictions were shown to be well-calibrated and sharp.

Adding solar X-ray flux data to empirical or machine-learned models can add useful information about transient solar activity, improving the 3-hour ahead prediction of the $\text{K}_\text{p}$ index for high geomagnetic activity levels.
Although including this relatively simple data set increased the accuracy of the forecast, supporting the suggestion that X-ray data is a reasonable proxy for solar magnetic activity, our model still shows lags in predicting large geomagnetic storms.
Capturing uncertainty, providing probabilistic predictions and improving our ability to capture transient behavior are all within reach with modern tools and do not require sacrificing model predictive performance.
We hope that future work continues to bring together recent advances in feature selection \citep[e.g.,][]{Zhelavskaya2019}, model design to accommodate probabilistic prediction, and more complex solar data sources such as solar magnetograms, to provide accurate forecasting of strong geomagnetic activity with longer lead times.

\begin{acknowledgements}
      SC thanks to the Space Science and Applications group and the Center for Space and Earth Science (CSES) at Los Alamos National Laboratory for organizing and supporting the Los Alamos Space Weather Summer School.
      Portions of this work by SKM were performed under the auspices of the US Department of Energy and were partially supported by the Laboratory Directed Research and Development (LDRD) program, award number 20190262ER.
      The authors wish to acknowledge the use of the OMNI solar wind data, available at \url{https://omniweb.gsfc.nasa.gov/ow.html}.
      The $\text{K}_\text{p}$ index and GOES X-ray datasets were accessed through the GFZ-Potsdam website \url{https://www.gfz-potsdam.de/en/kp-index/} and NOAA FTP server \url{https://satdat.ngdc.noaa.gov/sem/goes/data/}, respectively.
      The majority of analysis and visualization was completed with the help of free, open-source software tools, notably: Keras \citep{chollet2015keras}; matplotlib \citep{Hunter2007}; IPython \citep{PerezNoTitle}; pandas \citep{McKinney2010}; Spacepy \citep{spacepy11}; PyForecastTools \citep{PyForecastTools}; and others \citep[see, e.g.,][]{Millman2011}.
      The code developed during this work is available at \url{https://github.com/shibaji7/Codebase_Kp_Prediction}.
\end{acknowledgements}


\bibliography{references}

\begin{thebibliography}{80}
\providecommand{\natexlab}[1]{#1}
\providecommand{\url}[1]{\texttt{#1}}
\providecommand{\urlprefix}{URL }
\providecommand{\eprint}[2][]{\url{#2}}

\bibitem[{Afshartous and Preston(2010)}]{afshartous10}
Afshartous, D., and R.~A. Preston, 2010.
\newblock Confidence intervals for dependent data: Equating non-overlap with
  statistical significance.
\newblock \emph{Computational Statistics \& Data Analysis}, \textbf{54}(10),
  2296--2305.
\newblock 10.1016/j.csda.2010.04.011,
  \urlprefix\url{http://www.sciencedirect.com/science/article/pii/S0167947310001568}.

\bibitem[{Al-Shedivat et~al.(2017)Al-Shedivat, Wilson, Saatchi, Hu, and
  Xing}]{alshedivat2016srk}
Al-Shedivat, M., A.~G. Wilson, Y.~Saatchi, Z.~Hu, and E.~P. Xing, 2017.
\newblock {Learning Scalable Deep Kernels with Recurrent Structure}.
\newblock \emph{Journal of Machine Learning Research}, \textbf{18}(82), 1--37.
\newblock \urlprefix\url{http://jmlr.org/papers/v18/16-498.html}.

\bibitem[{{Arge} et~al.(2010){Arge}, {Henney}, {Koller}, {Compeau}, {Young},
  {MacKenzie}, {Fay}, and {Harvey}}]{Arge2010}
{Arge}, C.~N., C.~J. {Henney}, J.~{Koller}, C.~R. {Compeau}, S.~{Young},
  D.~{MacKenzie}, A.~{Fay}, and J.~W. {Harvey}, 2010.
\newblock {Air Force Data Assimilative Photospheric Flux Transport (ADAPT)
  Model}.
\newblock \emph{Twelfth International Solar Wind Conference}, \textbf{1216},
  343--346.
\newblock 10.1063/1.3395870,
  \urlprefix\url{https://aip.scitation.org/doi/abs/10.1063/1.3395870}.

\bibitem[{Baker et~al.(1981)Baker, Hones, Payne, and Feldman}]{Baker1981}
Baker, D.~N., E.~W. Hones, J.~B. Payne, and W.~C. Feldman, 1981.
\newblock A high time resolution study of interplanetary parameter correlations
  with AE.
\newblock \emph{Geophysical Research Letters}, \textbf{8}(2), 179--182.
\newblock 10.1029/GL008i002p00179,
  \urlprefix\url{https://agupubs.onlinelibrary.wiley.com/doi/abs/10.1029/GL008i002p00179}.

\bibitem[{Bala and Reiff(2012)}]{bala_improvements_2012}
Bala, R., and P.~Reiff, 2012.
\newblock {Improvements in short-term forecasting of geomagnetic activity}.
\newblock \emph{Space Weather}, \textbf{10}(6).
\newblock 10.1029/2012SW000779,
  \urlprefix\url{https://agupubs.onlinelibrary.wiley.com/doi/abs/10.1029/2012SW000779}.

\bibitem[{Bartels(1949)}]{Bartels1949}
Bartels, J.~R., 1949.
\newblock The standardized index, Ks and the planetary index, Kp.
\newblock \emph{IATME}, \textbf{97}(12b).

\bibitem[{{Bingham, Suzy} et~al.(2019){Bingham, Suzy}, {Murray, Sophie A.},
  {Guerrero, Antonio}, {Glover, Alexi}, and {Thorn, Peter}}]{Bingham2019}
{Bingham, Suzy}, {Murray, Sophie A.}, {Guerrero, Antonio}, {Glover, Alexi}, and
  {Thorn, Peter}, 2019.
\newblock Summary of the plenary sessions at European Space Weather Week 15:
  space weather users and service providers working together now and in the
  future.
\newblock \emph{J. Space Weather Space Clim.}, \textbf{9}, A32.
\newblock 10.1051/swsc/2019031,
  \urlprefix\url{https://doi.org/10.1051/swsc/2019031}.

\bibitem[{Boberg et~al.(2000)Boberg, Wintoft, and Lundstedt}]{boberg_real_2000}
Boberg, F., P.~Wintoft, and H.~Lundstedt, 2000.
\newblock {Real time {\{}Kp{\}} predictions from solar wind data using neural
  networks}.
\newblock \emph{Physics and Chemistry of the Earth, Part C: Solar, Terrestrial
  {\&} Planetary Science}, \textbf{25}(4), 275--280.
\newblock 10.1016/S1464-1917(00)00016-7,
  \urlprefix\url{http://www.sciencedirect.com/science/article/pii/S1464191700000167}.

\bibitem[{Borovsky(2013)}]{Borovsky2013}
Borovsky, J.~E., 2013.
\newblock Physical improvements to the solar wind reconnection control function
  for the Earth's magnetosphere.
\newblock \emph{Journal of Geophysical Research: Space Physics},
  \textbf{118}(5), 2113--2121.
\newblock 10.1002/jgra.50110,
  \urlprefix\url{https://agupubs.onlinelibrary.wiley.com/doi/abs/10.1002/jgra.50110}.

\bibitem[{Borovsky(2014)}]{Borovsky_canonical_2014}
Borovsky, J.~E., 2014.
\newblock Canonical correlation analysis of the combined solar wind and
  geomagnetic index data sets.
\newblock \emph{Journal of Geophysical Research: Space Physics},
  \textbf{119}(7), 5364--5381.
\newblock 10.1002/2013JA019607,
  \urlprefix\url{https://agupubs.onlinelibrary.wiley.com/doi/abs/10.1002/2013JA019607}.

\bibitem[{Borovsky and Denton(2006)}]{borovsky06}
Borovsky, J.~E., and M.~H. Denton, 2006.
\newblock Differences between CME-driven storms and CIR-driven storms.
\newblock \emph{Journal of Geophysical Research: Space Physics},
  \textbf{111}(A7).
\newblock 10.1029/2005JA011447,
  \urlprefix\url{https://agupubs.onlinelibrary.wiley.com/doi/abs/10.1029/2005JA011447}.

\bibitem[{Borovsky et~al.(1998)Borovsky, Thomsen, Elphic, Cayton, and
  McComas}]{borovsky_transport_1998}
Borovsky, J.~E., M.~F. Thomsen, R.~C. Elphic, T.~E. Cayton, and D.~J. McComas,
  1998.
\newblock {The transport of plasma sheet material from the distant tail to
  geosynchronous orbit}.
\newblock \emph{Journal of Geophysical Research: Space Physics},
  \textbf{103}(A9), 20,297--20,331.
\newblock 10.1029/97JA03144,
  \urlprefix\url{https://agupubs.onlinelibrary.wiley.com/doi/abs/10.1029/97JA03144}.

\bibitem[{Bradley(1997)}]{Bradley1997}
Bradley, A.~P., 1997.
\newblock The use of the area under the ROC curve in the evaluation of machine
  learning algorithms.
\newblock \emph{Pattern Recognition}, \textbf{30}(7), 1145--1159.
\newblock 10.1016/S0031-3203(96)00142-2,
  \urlprefix\url{http://www.sciencedirect.com/science/article/pii/S0031320396001422}.

\bibitem[{Camporeale(2019)}]{Camporeale2019}
Camporeale, E., 2019.
\newblock The Challenge of Machine Learning in Space Weather: Nowcasting and
  Forecasting.
\newblock \emph{Space Weather}, \textbf{17}(8), 1166--1207.
\newblock 10.1029/2018SW002061,
  \urlprefix\url{https://agupubs.onlinelibrary.wiley.com/doi/abs/10.1029/2018SW002061}.

\bibitem[{Carbary(2005)}]{Carbary2005ABoundaries}
Carbary, J.~F., 2005.
\newblock {A Kp--based model of auroral boundaries}.
\newblock \emph{Space Weather}, \textbf{3}(10).
\newblock 10.1029/2005SW000162,
  \urlprefix\url{http://doi.wiley.com/10.1029/2005SW000162}.

\bibitem[{Choi et~al.(2011)Choi, Lee, Cho, Kwak, Cho, Park, Kim, Baker, Reeves,
  and Lee}]{Choi2011}
Choi, H.-S., J.~Lee, K.-S. Cho, Y.-S. Kwak, I.-H. Cho, Y.-D. Park, Y.-H. Kim,
  D.~N. Baker, G.~D. Reeves, and D.-K. Lee, 2011.
\newblock Analysis of GEO spacecraft anomalies: Space weather relationships.
\newblock \emph{Space Weather}, \textbf{9}(6).
\newblock 10.1029/2010SW000597,
  \urlprefix\url{https://agupubs.onlinelibrary.wiley.com/doi/abs/10.1029/2010SW000597}.

\bibitem[{Chollet et~al.(2015)}]{chollet2015keras}
Chollet, F., et~al., 2015.
\newblock Keras.
\newblock \url{https://keras.io}.

\bibitem[{Costello(1998)}]{costello_moving_1998}
Costello, K.~A., 1998.
\newblock {Moving the {Rice} {MSFM} into a real-time forecast mode using solar
  wind driven forecast modules}.
\newblock Ph.D. thesis.
\newblock \urlprefix\url{https://scholarship.rice.edu/handle/1911/19251}.

\bibitem[{DeLong et~al.(1988)DeLong, DeLong, and Clarke-Pearson}]{delong88}
DeLong, E.~R., D.~M. DeLong, and D.~L. Clarke-Pearson, 1988.
\newblock Comparing the Areas under Two or More Correlated Receiver Operating
  Characteristic Curves: A Nonparametric Approach.
\newblock \emph{Biometrics}, \textbf{44}(3), 837--845.
\newblock 10.2307/2531595, \urlprefix\url{http://www.jstor.org/stable/2531595}.

\bibitem[{Dungey(1961)}]{Dungey1961}
Dungey, J.~W., 1961.
\newblock {Interplanetary Magnetic Field and the Auroral Zones}.
\newblock \emph{Phys. Rev. Lett.}, \textbf{6}(2), 47--48.
\newblock 10.1103/PhysRevLett.6.47,
  \urlprefix\url{https://link.aps.org/doi/10.1103/PhysRevLett.6.47}.

\bibitem[{Eastwood et~al.(2017)Eastwood, Biffis, Hapgood, Green, Bisi, Bentley,
  Wicks, McKinnell, Gibbs, and Burnett}]{Eastwood2017}
Eastwood, J.~P., E.~Biffis, M.~A. Hapgood, L.~Green, M.~M. Bisi, R.~D. Bentley,
  R.~Wicks, L.-A. McKinnell, M.~Gibbs, and C.~Burnett, 2017.
\newblock The Economic Impact of Space Weather: Where Do We Stand?
\newblock \emph{Risk Analysis}, \textbf{37}(2), 206--218.
\newblock 10.1111/risa.12765,
  \urlprefix\url{https://onlinelibrary.wiley.com/doi/abs/10.1111/risa.12765}.

\bibitem[{Estabrooks et~al.(2004)Estabrooks, Jo, and
  Japkowicz}]{Estabrooks2004}
Estabrooks, A., T.~Jo, and N.~Japkowicz, 2004.
\newblock {A Multiple Resampling Method for Learning from Imbalanced Data
  Sets}.
\newblock \emph{Computational Intelligence}, \textbf{20}(1), 18--36.
\newblock 10.1111/j.0824-7935.2004.t01-1-00228.x,
  \urlprefix\url{https://onlinelibrary.wiley.com/doi/abs/10.1111/j.0824-7935.2004.t01-1-00228.x}.

\bibitem[{Garcia(1994)}]{garcia1994temperature}
Garcia, H.~A., 1994.
\newblock Temperature and emission measure from GOES soft X-ray measurements.
\newblock \emph{Solar Physics}, \textbf{154}(2), 275--308.
\newblock 10.1007/BF00681100,
  \urlprefix\url{https://link.springer.com/article/10.1007/BF00681100}.

\bibitem[{Gonzalez et~al.(1999)Gonzalez, Tsurutani, and Cl{\'u}a~de
  Gonzalez}]{Gonzalez1999}
Gonzalez, W.~D., B.~T. Tsurutani, and A.~L. Cl{\'u}a~de Gonzalez, 1999.
\newblock Interplanetary origin of geomagnetic storms.
\newblock \emph{Space Science Reviews}, \textbf{88}(3), 529--562.
\newblock 10.1023/A:1005160129098,
  \urlprefix\url{https://doi.org/10.1023/A:1005160129098}.

\bibitem[{Haiducek et~al.(2017)Haiducek, Welling, Ganushkina, Morley, and
  Ozturk}]{Haiducek2017}
Haiducek, J.~D., D.~T. Welling, N.~Y. Ganushkina, S.~K. Morley, and D.~S.
  Ozturk, 2017.
\newblock SWMF Global Magnetosphere Simulations of January 2005: Geomagnetic
  Indices and Cross-Polar Cap Potential.
\newblock \emph{Space Weather}, \textbf{15}(12), 1567--1587.
\newblock 10.1002/2017SW001695,
  \urlprefix\url{https://agupubs.onlinelibrary.wiley.com/doi/abs/10.1002/2017SW001695}.

\bibitem[{Hochreiter and Schmidhuber(1997)}]{Sepp1997}
Hochreiter, S., and J.~Schmidhuber, 1997.
\newblock Long Short-Term Memory.
\newblock \emph{Neural Comput.}, \textbf{9}(8), 1735–1780.
\newblock 10.1162/neco.1997.9.8.1735,
  \urlprefix\url{https://doi.org/10.1162/neco.1997.9.8.1735}.

\bibitem[{Horne et~al.(2013)Horne, Glauert, Meredith, Boscher, Maget,
  Heynderickx, and Pitchford}]{Horne2013}
Horne, R.~B., S.~A. Glauert, N.~P. Meredith, D.~Boscher, V.~Maget,
  D.~Heynderickx, and D.~Pitchford, 2013.
\newblock Space weather impacts on satellites and forecasting the Earth's
  electron radiation belts with SPACECAST.
\newblock \emph{Space Weather}, \textbf{11}(4), 169--186.
\newblock 10.1002/swe.20023,
  \urlprefix\url{https://agupubs.onlinelibrary.wiley.com/doi/abs/10.1002/swe.20023}.

\bibitem[{Hundhausen(1970)}]{Hundhausen1970}
Hundhausen, A.~J., 1970.
\newblock Composition and dynamics of the solar wind plasma.
\newblock \emph{Reviews of Geophysics}, \textbf{8}(4), 729--811.
\newblock 10.1029/RG008i004p00729,
  \urlprefix\url{https://agupubs.onlinelibrary.wiley.com/doi/abs/10.1029/RG008i004p00729}.

\bibitem[{Hunter(2007)}]{Hunter2007}
Hunter, J.~D., 2007.
\newblock {Matplotlib: A 2D graphics environment}.
\newblock \emph{Computing In Science {\&} Engineering}, \textbf{9}(3), 90--95.
\newblock 10.1109/MCSE.2007.55,
  \urlprefix\url{https://ieeexplore.ieee.org/document/4160265}.

\bibitem[{Johnson et~al.(2018)Johnson, Wing, and Camporeale}]{Johnson2018}
Johnson, J.~R., S.~Wing, and E.~Camporeale, 2018.
\newblock Transfer entropy and cumulant-based cost as measures of nonlinear
  causal relationships in space plasmas: applications to $D_\mathrm{st}$.
\newblock \emph{Annales Geophysicae}, \textbf{36}(4), 945--952.
\newblock 10.5194/angeo-36-945-2018,
  \urlprefix\url{https://www.ann-geophys.net/36/945/2018/}.

\bibitem[{Kahler and Ling(2018)}]{Kahler2018ForecastingRatios}
Kahler, S.~W., and A.~G. Ling, 2018.
\newblock {Forecasting Solar Energetic Particle (SEP) events with Flare X-ray
  peak ratios}.
\newblock \emph{Journal of Space Weather and Space Climate}, \textbf{8}, A47.
\newblock 10.1051/swsc/2018033,
  \urlprefix\url{https://www.swsc-journal.org/10.1051/swsc/2018033}.

\bibitem[{Kay et~al.(2003)Kay, Harra, Matthews, Culhane, and
  Green}]{Kay2003TheCMEs}
Kay, H. R.~M., L.~K. Harra, S.~A. Matthews, J.~L. Culhane, and L.~M. Green,
  2003.
\newblock {The soft X-ray characteristics of solar flares, both with and
  without associated CMEs}.
\newblock \emph{Astronomy {\&} Astrophysics}, \textbf{400}(2), 779--784.
\newblock 10.1051/0004-6361:20030095,
  \urlprefix\url{http://www.aanda.org/10.1051/0004-6361:20030095}.

\bibitem[{Kilpua et~al.(2009)Kilpua, Luhmann, Gosling, Li, Elliott
  et~al.}]{kilpua_small_2009}
Kilpua, E. K.~J., J.~G. Luhmann, J.~Gosling, Y.~Li, H.~Elliott, et~al., 2009.
\newblock Small Solar Wind Transients and Their Connection to the Large-Scale
  Coronal Structure.
\newblock \emph{Solar Physics}, \textbf{256}(1), 327--344.
\newblock 10.1007/s11207-009-9366-1,
  \urlprefix\url{https://doi.org/10.1007/s11207-009-9366-1}.

\bibitem[{Liemohn et~al.(2018)Liemohn, McCollough, Jordanova, Ngwira, Morley
  et~al.}]{Liemohn2018ModelPredictions}
Liemohn, M.~W., J.~P. McCollough, V.~K. Jordanova, C.~M. Ngwira, S.~K. Morley,
  et~al., 2018.
\newblock {Model Evaluation Guidelines for Geomagnetic Index Predictions}.
\newblock \emph{Space Weather}, \textbf{16}(12), 2079--2102.
\newblock 10.1029/2018SW002067,
  \urlprefix\url{http://doi.wiley.com/10.1029/2018SW002067}.

\bibitem[{Lippiello et~al.(2008)Lippiello, de~Arcangelis, and
  Godano}]{Lippiello2008DifferentEjections}
Lippiello, E., L.~de~Arcangelis, and C.~Godano, 2008.
\newblock {Different triggering mechanisms for solar flares and coronal mass
  ejections}.
\newblock \emph{Astronomy {\&} Astrophysics}, \textbf{488}(2), L29--L32.
\newblock 10.1051/0004-6361:200810164,
  \urlprefix\url{http://www.aanda.org/10.1051/0004-6361:200810164}.

\bibitem[{Luo et~al.(2017)Luo, Liu, and Gong}]{luo_two_2017}
Luo, B., S.~Liu, and J.~Gong, 2017.
\newblock {Two empirical models for short-term forecast of Kp}.
\newblock \emph{Space Weather}, \textbf{15}(3), 503--516.
\newblock 10.1002/2016SW001585,
  \urlprefix\url{https://agupubs.onlinelibrary.wiley.com/doi/abs/10.1002/2016SW001585}.

\bibitem[{Mayaud(1980)}]{Mayaud1980}
Mayaud, P.~N., 1980.
\newblock Derivation, Meaning and Use of Geomagnetic Indices, vol.~22 of
  \emph{Geophysical Monograph}.
\newblock American Geophysical Union.
\newblock 10.1029/GM022,
  \urlprefix\url{https://agupubs.onlinelibrary.wiley.com/doi/book/10.1029/GM022}.

\bibitem[{McKinney(2010)}]{McKinney2010}
McKinney, W., 2010.
\newblock {Data Structures for Statistical Computing in Python}.
\newblock In S.~van~der Walt and J.~Millman, eds., Proceedings of the 9th
  Python in Science Conference, 56--61.
\newblock 10.25080/Majora-92bf1922-012,
  \urlprefix\url{https://conference.scipy.org/proceedings/scipy2010/mckinney.html}.

\bibitem[{Michalek(2009)}]{Michalek2009TwoEjections}
Michalek, G., 2009.
\newblock {Two types of flare-associated coronal mass ejections}.
\newblock \emph{Astronomy {\&} Astrophysics}, \textbf{494}(1), 263--268.
\newblock 10.1051/0004-6361:200810662,
  \urlprefix\url{http://www.aanda.org/10.1051/0004-6361:200810662}.

\bibitem[{Millman and Aivazis(2011)}]{Millman2011}
Millman, K.~J., and M.~Aivazis, 2011.
\newblock {Python for Scientists and Engineers}.
\newblock \emph{Computing in Science {\&} Engineering}, \textbf{13}(2), 9--12.
\newblock 10.1109/MCSE.2011.36,
  \urlprefix\url{https://ieeexplore.ieee.org/document/5725235}.

\bibitem[{Morley(2018)}]{PyForecastTools}
Morley, S., 2018.
\newblock {drsteve/PyForecastTools: PyForecastTools}.
\newblock 10.5281/zenodo.1256921,
  \urlprefix\url{https://doi.org/10.5281/zenodo.1256921}.

\bibitem[{Morley(2019)}]{Morley2019}
Morley, S.~K., 2019.
\newblock Challenges and opportunities in magnetospheric space weather
  prediction.
\newblock \emph{Space Weather}, \textbf{n/a}(n/a).
\newblock 10.1029/2018SW002108,
  \eprint{https://agupubs.onlinelibrary.wiley.com/doi/pdf/10.1029/2018SW002108}.

\bibitem[{Morley et~al.(2018{\natexlab{a}})Morley, Brito, and
  Welling}]{MorleyMetrics2018}
Morley, S.~K., T.~V. Brito, and D.~T. Welling, 2018{\natexlab{a}}.
\newblock Measures of Model Performance Based On the Log Accuracy Ratio.
\newblock \emph{Space Weather}, \textbf{16}(1), 69--88.
\newblock 10.1002/2017SW001669,
  \urlprefix\url{https://agupubs.onlinelibrary.wiley.com/doi/abs/10.1002/2017SW001669}.

\bibitem[{Morley et~al.(2011)Morley, Koller, Welling, Larsen, Henderson, and
  Niehof}]{spacepy11}
Morley, S.~K., J.~Koller, D.~T. Welling, B.~A. Larsen, M.~G. Henderson, and
  J.~T. Niehof, 2011.
\newblock {Spacepy - A Python-based library of tools for the space sciences}.
\newblock In Proceedings of the 9th Python in science conference (SciPy 2010),
  67--71. Austin, TX.
\newblock
  \urlprefix\url{https://conference.scipy.org/proceedings/scipy2010/pdfs/morley.pdf}.

\bibitem[{Morley et~al.(2018{\natexlab{b}})Morley, Welling, and
  Woodroffe}]{morley18swmf}
Morley, S.~K., D.~T. Welling, and J.~R. Woodroffe, 2018{\natexlab{b}}.
\newblock Perturbed Input Ensemble Modeling With the Space Weather Modeling
  Framework.
\newblock \emph{Space Weather}, \textbf{16}(9), 1330--1347.
\newblock 10.1029/2018SW002000,
  \urlprefix\url{https://agupubs.onlinelibrary.wiley.com/doi/abs/10.1029/2018SW002000}.

\bibitem[{Murphy(1995)}]{Murphy1995}
Murphy, A.~H., 1995.
\newblock The Coefficients of Correlation and Determination as Measures of
  performance in Forecast Verification.
\newblock \emph{Weather and Forecasting}, \textbf{10}(4), 681--688.
\newblock 10.1175/1520-0434(1995)010$<$0681:TCOCAD$>$2.0.CO;2,
  \urlprefix\url{https://doi.org/10.1175/1520-0434(1995)010<0681:TCOCAD>2.0.CO;2}.

\bibitem[{Newell et~al.(2007)Newell, Sotirelis, Liou, Meng, and
  Rich}]{newell_nearly_2007}
Newell, P.~T., T.~Sotirelis, K.~Liou, C.-I. Meng, and F.~J. Rich, 2007.
\newblock {A nearly universal solar wind-magnetosphere coupling function
  inferred from 10 magnetospheric state variables}.
\newblock \emph{Journal of Geophysical Research: Space Physics},
  \textbf{112}(A1).
\newblock 10.1029/2006JA012015,
  \urlprefix\url{https://agupubs.onlinelibrary.wiley.com/doi/abs/10.1029/2006JA012015}.

\bibitem[{N{\'{u}}{\~{n}}ez(2018)}]{Nunez2018PredictingElectrons}
N{\'{u}}{\~{n}}ez, M., 2018.
\newblock {Predicting well-connected SEP events from observations of solar soft
  X-rays and near-relativistic electrons}.
\newblock \emph{Journal of Space Weather and Space Climate}, \textbf{8}, A36.
\newblock 10.1051/swsc/2018023,
  \urlprefix\url{https://www.swsc-journal.org/10.1051/swsc/2018023}.

\bibitem[{O'Brien(2009)}]{OBrien2009}
O'Brien, T.~P., 2009.
\newblock SEAES-GEO: A spacecraft environmental anomalies expert system for
  geosynchronous orbit.
\newblock \emph{Space Weather}, \textbf{7}(9).
\newblock 10.1029/2009SW000473,
  \urlprefix\url{https://agupubs.onlinelibrary.wiley.com/doi/abs/10.1029/2009SW000473}.

\bibitem[{Osthus et~al.(2014)Osthus, Caragea, Higdon, Morley, Reeves, and
  Weaver}]{osthus14}
Osthus, D., P.~C. Caragea, D.~Higdon, S.~K. Morley, G.~D. Reeves, and B.~P.
  Weaver, 2014.
\newblock Dynamic linear models for forecasting of radiation belt electrons and
  limitations on physical interpretation of predictive models.
\newblock \emph{Space Weather}, \textbf{12}(6), 426--446.
\newblock 10.1002/2014SW001057,
  \urlprefix\url{https://agupubs.onlinelibrary.wiley.com/doi/abs/10.1002/2014SW001057}.

\bibitem[{{Perez} and {Granger}(2007)}]{PerezNoTitle}
{Perez}, F., and B.~E. {Granger}, 2007.
\newblock IPython: A System for Interactive Scientific Computing.
\newblock \emph{Computing in Science Engineering}, \textbf{9}(3), 21--29.
\newblock 10.1109/MCSE.2007.53,
  \urlprefix\url{https://ieeexplore.ieee.org/document/4160251}.

\bibitem[{{Qiu} et~al.(2015){Qiu}, {Fleeman}, and {Ball}}]{Qiu2015}
{Qiu}, Q., J.~A. {Fleeman}, and D.~R. {Ball}, 2015.
\newblock Geomagnetic Disturbance: A comprehensive approach by American
  Electric Power to address the impacts.
\newblock \emph{IEEE Electrification Magazine}, \textbf{3}(4), 22--33.
\newblock 10.1109/MELE.2015.2480615,
  \urlprefix\url{https://ieeexplore.ieee.org/abstract/document/7343043}.

\bibitem[{Rasmussen and Williams(2006)}]{Rasmussen2006GaussianLearning}
Rasmussen, C.~E., and C.~K.~I. Williams, 2006.
\newblock {Gaussian processes for machine learning}.
\newblock MIT Press.
\newblock ISBN 026218253X.
\newblock \urlprefix\url{http://www.gaussianprocess.org/gpml/}.

\bibitem[{Revelle(2020)}]{revelle20}
Revelle, W.~R., 2020.
\newblock {psych: Procedures for Personality and Psychological Research}.
\newblock R package version 1.9.12.31,
  \urlprefix\url{https://CRAN.R-project.org/package=psych}.

\bibitem[{Richardson and Cane(2012)}]{Richardson2012}
Richardson, I.~G., and H.~V. Cane, 2012.
\newblock Solar wind drivers of geomagnetic storms during more than four solar
  cycles.
\newblock \emph{J. Space Weather Space Clim.}, \textbf{2}, A01.
\newblock 10.1051/swsc/2012001,
  \urlprefix\url{https://doi.org/10.1051/swsc/2012001}.

\bibitem[{Robin et~al.(2011)Robin, Turck, Hainard, Tiberti, Lisacek, Sanchez,
  and M\:{u}ller}]{robin14}
Robin, X., N.~Turck, A.~Hainard, N.~Tiberti, F.~Lisacek, J.~C. Sanchez, and
  M.~M\:{u}ller, 2011.
\newblock {pROC}: an open-source package for {R} and {S+} to analyze and
  compare {ROC} curves.
\newblock \emph{BMC Bioinformatics}, \textbf{12}(77).
\newblock 10.1186/1471-2105-12-77,
  \urlprefix\url{https://bmcbioinformatics.biomedcentral.com/articles/10.1186/1471-2105-12-77}.

\bibitem[{Schrijver et~al.(2015)Schrijver, Kauristie, Aylward, Denardini,
  Gibson et~al.}]{Schrijver2015}
Schrijver, C.~J., K.~Kauristie, A.~D. Aylward, C.~M. Denardini, S.~E. Gibson,
  et~al., 2015.
\newblock Understanding space weather to shield society: A global road map for
  2015-2025 commissioned by COSPAR and ILWS.
\newblock \emph{Advances in Space Research}, \textbf{55}(12), 2745 -- 2807.
\newblock 10.1016/j.asr.2015.03.023,
  \urlprefix\url{http://www.sciencedirect.com/science/article/pii/S0273117715002252}.

\bibitem[{Schwenn et~al.(2005)Schwenn, Dal~Lago, Huttunen, and
  Gonzalez}]{Schwenn2005}
Schwenn, R., A.~Dal~Lago, E.~Huttunen, and W.~D. Gonzalez, 2005.
\newblock The association of coronal mass ejections with their effects near the
  Earth.
\newblock \emph{Annales Geophysicae}, \textbf{23}(3), 1033--1059.
\newblock 10.5194/angeo-23-1033-2005,
  \urlprefix\url{https://angeo.copernicus.org/articles/23/1033/2005/}.

\bibitem[{Sexton et~al.(2019)Sexton, Nykyri, and Ma}]{Sexton2019KpNetwork}
Sexton, E.~S., K.~Nykyri, and X.~Ma, 2019.
\newblock {Kp forecasting with a recurrent neural network}.
\newblock \emph{Journal of Space Weather and Space Climate}, \textbf{9}, A19.
\newblock 10.1051/swsc/2019020,
  \urlprefix\url{https://www.swsc-journal.org/10.1051/swsc/2019020}.

\bibitem[{Sharpe and Murray(2017)}]{Sharpe2017}
Sharpe, M.~A., and S.~A. Murray, 2017.
\newblock Verification of Space Weather Forecasts Issued by the Met Office
  Space Weather Operations Centre.
\newblock \emph{Space Weather}, \textbf{15}(10), 1383--1395.
\newblock 10.1002/2017SW001683,
  \urlprefix\url{https://agupubs.onlinelibrary.wiley.com/doi/abs/10.1002/2017SW001683}.

\bibitem[{Shprits et~al.(2019)Shprits, Vasile, and Zhelavskaya}]{Shprits2019}
Shprits, Y.~Y., R.~Vasile, and I.~S. Zhelavskaya, 2019.
\newblock Nowcasting and Predicting the Kp Index Using Historical Values and
  Real-Time Observations.
\newblock \emph{Space Weather}, \textbf{17}(8), 1219--1229.
\newblock 10.1029/2018SW002141,
  \urlprefix\url{https://agupubs.onlinelibrary.wiley.com/doi/abs/10.1029/2018SW002141}.

\bibitem[{Srivastava and Venkatakrishnan(2002)}]{Srivastava2002}
Srivastava, N., and P.~Venkatakrishnan, 2002.
\newblock Relationship between CME Speed and Geomagnetic Storm Intensity.
\newblock \emph{Geophysical Research Letters}, \textbf{29}(9).
\newblock 10.1029/2001GL013597,
  \urlprefix\url{https://agupubs.onlinelibrary.wiley.com/doi/abs/10.1029/2001GL013597}.

\bibitem[{Srivastava and Venkatakrishnan(2004)}]{Srivastava2004}
Srivastava, N., and P.~Venkatakrishnan, 2004.
\newblock Solar and interplanetary sources of major geomagnetic storms during
  1996–2002.
\newblock \emph{Journal of Geophysical Research: Space Physics},
  \textbf{109}(A10).
\newblock 10.1029/2003JA010175,
  \urlprefix\url{https://agupubs.onlinelibrary.wiley.com/doi/abs/10.1029/2003JA010175}.

\bibitem[{Steiger(1980)}]{steiger80}
Steiger, J.~H., 1980.
\newblock Tests for Comparing Elements of a Correlation Matrix.
\newblock \emph{Psychological Bulletin}, \textbf{87}(2), 245--251.
\newblock 10.1037/0033-2909.87.2.245,
  \urlprefix\url{https://psycnet.apa.org/record/1980-08757-001}.

\bibitem[{{Sun} and {Xu}(2014)}]{Sun14}
{Sun}, X., and W.~{Xu}, 2014.
\newblock Fast Implementation of DeLong’s Algorithm for Comparing the Areas
  Under Correlated Receiver Operating Characteristic Curves.
\newblock \emph{IEEE Signal Processing Letters}, \textbf{21}(11), 1389--1393.
\newblock 10.1109/LSP.2014.2337313,
  \urlprefix\url{https://ieeexplore.ieee.org/document/6851192}.

\bibitem[{Tan et~al.(2018)Tan, Hu, Wang, and Zhong}]{tan_geomagnetic_2018}
Tan, Y., Q.~Hu, Z.~Wang, and Q.~Zhong, 2018.
\newblock {Geomagnetic Index Kp Forecasting With LSTM}.
\newblock \emph{Space Weather}, \textbf{16}(4), 406--416.
\newblock 10.1002/2017SW001764,
  \urlprefix\url{https://agupubs.onlinelibrary.wiley.com/doi/abs/10.1002/2017SW001764}.

\bibitem[{T\'{o}th et~al.(2005)T\'{o}th, Sokolov, Gombosi, Chesney, Clauer
  et~al.}]{Toth2005}
T\'{o}th, G., I.~V. Sokolov, T.~I. Gombosi, D.~R. Chesney, C.~R. Clauer,
  et~al., 2005.
\newblock Space Weather Modeling Framework: A new tool for the space science
  community.
\newblock \emph{Journal of Geophysical Research: Space Physics},
  \textbf{110}(A12).
\newblock 10.1029/2005JA011126,
  \urlprefix\url{https://agupubs.onlinelibrary.wiley.com/doi/abs/10.1029/2005JA011126}.

\bibitem[{Wilks(2006)}]{wilks06}
Wilks, D.~S., 2006.
\newblock Statistical methods in the atmospheric sciences, 2nd Edition.
\newblock Academic Press.
\newblock ISBN 9780123850232.
\newblock
  \urlprefix\url{https://www.elsevier.com/books/statistical-methods-in-the-atmospheric-sciences/wilks/978-0-12-385022-5}.

\bibitem[{Wing et~al.(2016)Wing, Johnson, Camporeale, and
  Reeves}]{Wing2016InformationBelt}
Wing, S., J.~R. Johnson, E.~Camporeale, and G.~D. Reeves, 2016.
\newblock {Information theoretical approach to discovering solar wind drivers
  of the outer radiation belt}.
\newblock \emph{Journal of Geophysical Research: Space Physics},
  \textbf{121}(10), 9378--9399.
\newblock 10.1002/2016JA022711,
  \urlprefix\url{http://doi.wiley.com/10.1002/2016JA022711}.

\bibitem[{Wing et~al.(2005)Wing, Johnson, Jen, Meng, Sibeck, Bechtold, Freeman,
  Costello, Balikhin, and Takahashi}]{wing_kp_2005}
Wing, S., J.~R. Johnson, J.~Jen, C.-I. Meng, D.~G. Sibeck, K.~Bechtold,
  J.~Freeman, K.~Costello, M.~Balikhin, and K.~Takahashi, 2005.
\newblock {Kp forecast models}.
\newblock \emph{Journal of Geophysical Research: Space Physics},
  \textbf{110}(A4).
\newblock 10.1029/2004JA010500,
  \urlprefix\url{https://agupubs.onlinelibrary.wiley.com/doi/abs/10.1029/2004JA010500}.

\bibitem[{Winter and Balasubramaniam(2015)}]{Winter2015UsingClass}
Winter, L.~M., and K.~Balasubramaniam, 2015.
\newblock {Using the maximum X-ray flux ratio and X-ray background to predict
  solar flare class}.
\newblock \emph{Space Weather}, \textbf{13}(5), 286--297.
\newblock 10.1002/2015SW001170,
  \urlprefix\url{http://doi.wiley.com/10.1002/2015SW001170}.

\bibitem[{Winter and Balasubramaniam(2014)}]{winter_estimate_2014-1}
Winter, L.~M., and K.~S. Balasubramaniam, 2014.
\newblock {Estimate of Solar Maximum using the 1--8\AA { } Geostationary
  Operational Environmental Satellites X--ray Measurements}.
\newblock \emph{The Astrophysical Journal}, \textbf{793}(2), L45.
\newblock 10.1088/2041-8205/793/2/l45,
  \urlprefix\url{https://iopscience.iop.org/article/10.1088/2041-8205/793/2/L45/pdf}.

\bibitem[{Wintoft et~al.(2017)Wintoft, Wik, Matzka, and
  Shprits}]{Wintoft2017ForecastingValues}
Wintoft, P., M.~Wik, J.~Matzka, and Y.~Shprits, 2017.
\newblock {Forecasting Kp from solar wind data: input parameter study using
  3-hour averages and 3-hour range values}.
\newblock \emph{Journal of Space Weather and Space Climate}, \textbf{7}, A29.
\newblock 10.1051/SWSC/2017027,
  \urlprefix\url{https://www.swsc-journal.org/articles/swsc/abs/2017/01/swsc160051/swsc160051.html}.

\bibitem[{Wu et~al.(2008)Wu, Feng, and Chao}]{Wu2008EnergyActivity}
Wu, D.~J., H.~Q. Feng, and J.~K. Chao, 2008.
\newblock {Energy spectrum of interplanetary magnetic flux ropes and its
  connection with solar activity}.
\newblock \emph{Astronomy {\&} Astrophysics}, \textbf{480}(1), L9--L12.
\newblock 10.1051/0004-6361:20079173,
  \urlprefix\url{http://www.aanda.org/10.1051/0004-6361:20079173}.

\bibitem[{Xu and Borovsky(2015{\natexlab{a}})}]{xu_new_2015}
Xu, F., and J.~E. Borovsky, 2015{\natexlab{a}}.
\newblock {A new four-plasma categorization scheme for the solar wind}.
\newblock \emph{Journal of Geophysical Research: Space Physics},
  \textbf{120}(1), 70--100.
\newblock 10.1002/2014JA020412,
  \urlprefix\url{https://agupubs.onlinelibrary.wiley.com/doi/abs/10.1002/2014JA020412}.

\bibitem[{Xu and Borovsky(2015{\natexlab{b}})}]{Borovsky2014}
Xu, F., and J.~E. Borovsky, 2015{\natexlab{b}}.
\newblock {A new four-plasma categorization scheme for the solar wind}.
\newblock \emph{Journal of Geophysical Research: Space Physics},
  \textbf{120}(1), 70--100.
\newblock 10.1002/2014JA020412,
  \urlprefix\url{https://agupubs.onlinelibrary.wiley.com/doi/abs/10.1002/2014JA020412}.

\bibitem[{Yap et~al.(2014)Yap, Rani, Rahman, Fong, Khairudin, and
  Abdullah}]{Yap2014}
Yap, B.~W., K.~A. Rani, H.~A.~A. Rahman, S.~Fong, Z.~Khairudin, and N.~N.
  Abdullah, 2014.
\newblock An Application of Oversampling, Undersampling, Bagging and Boosting
  in Handling Imbalanced Datasets.
\newblock In T.~Herawan, M.~M. Deris, and J.~Abawajy, eds., Proceedings of the
  First International Conference on Advanced Data and Information Engineering
  (DaEng-2013), 13--22. Springer Singapore, Singapore.
\newblock ISBN 978-981-4585-18-7.

\bibitem[{Zhang et~al.(2018)Zhang, Blanco-Cano, Nitta, Srivastava, and
  Mandrini}]{zhang18transients}
Zhang, J., X.~Blanco-Cano, N.~Nitta, N.~Srivastava, and C.~H. Mandrini, 2018.
\newblock Editorial: Earth-affecting Solar Transients.
\newblock \emph{Solar Physics}, \textbf{293}.
\newblock 10.1007/s11207-018-1302-9,
  \urlprefix\url{https://link.springer.com/article/10.1007/s11207-018-1302-9}.

\bibitem[{Zhelavskaya et~al.(2019)Zhelavskaya, Vasile, Shprits, Stolle, and
  Matzka}]{Zhelavskaya2019}
Zhelavskaya, I.~S., R.~Vasile, Y.~Y. Shprits, C.~Stolle, and J.~Matzka, 2019.
\newblock Systematic Analysis of Machine Learning and Feature Selection
  Techniques for Prediction of the Kp Index.
\newblock \emph{Space Weather}, \textbf{0}(ja).
\newblock 10.1029/2019SW002271,
  \urlprefix\url{https://agupubs.onlinelibrary.wiley.com/doi/abs/10.1029/2019SW002271}.

\bibitem[{Zhou et~al.(2003)Zhou, Wang, and Cao}]{zhou_correlation_2003}
Zhou, G., J.~Wang, and Z.~Cao, 2003.
\newblock {Correlation between halo coronal mass ejections and solar surface
  activity}.
\newblock \emph{A{\&}A}, \textbf{397}(3), 1057--1067.
\newblock 10.1051/0004-6361:20021463,
  \urlprefix\url{https://www.aanda.org/articles/aa/abs/2003/03/aa2720/aa2720.html}.

\end{thebibliography}

\end{document}